\documentclass[fleqn,usenatbib]{mnras}

\usepackage[T1]{fontenc}

\DeclareRobustCommand{\VAN}[3]{#2}
\let\VANthebibliography\thebibliography
\def\thebibliography{\DeclareRobustCommand{\VAN}[3]{##3}\VANthebibliography}

\usepackage{graphicx}	
\usepackage{amsmath}	
\usepackage{amssymb}	

\usepackage[normalem]{ulem}

\title[ZTF Year 1 Cosmology]{The Zwicky Transient Facility Type Ia supernova survey: first data release and results}

\author[S. Dhawan et al. ]{
S. Dhawan,$^{1,2}$\thanks{E-mail: suhail.dhawan@ast.cam.ac.uk} A. Goobar,$^{1}$ M.Smith,$^{3}$ J.Johansson,$^{1}$ M.Rigault,$^{3}$ J.Nordin,$^{4}$ R. Biswas,$^{1}$ D.Goldstein,$^{5}$
\newauthor
P. Nugent,$^{5,6}$ Y.-L. Kim,$^{3}$ A. A. Miller,$^{7,8}$ M.J. Graham,$^{9}$ M. Medford,$^{5,6}$ M. M. Kasliwal,$^{9}$ S. R. Kulkarni,$^{9}$ 
\newauthor
Dmitry A. Duev,$^{9}$ E. Bellm,$^{10}$ P. Rosnet,$^{11}$ R. Riddle,$^{12}$ J. Sollerman$^{13}$\\
$^{1}$The Oskar Klein Centre for Cosmoparticle Physics, Department of Physics, Stockholm University, SE-10691 Stockholm, Sweden\\
$^{2}$Kavli Institute for Cosmology and Institute of Astronomy, University of Cambridge, Madingley Road, Cambridge CB3 0HA, UK\\
$^{3}$Universit\'e de Lyon, Universit\'e Claude Bernard Lyon 1, CNRS/IN2P3, IP2I Lyon, F-69622, Villeurbanne, France.\\
$^{4}$Institute of Physics, Humboldt-Universität zu Berlin, Newtonstr. 15, 12489 Berlin, Germany\\
$^{5}$E.O. Lawrence Berkeley National Laboratory, 1 Cyclotron Rd., Berkeley, CA, 94720\\
$^{6}$Department of Astronomy, University of California, Berkeley, CA 94720-3411, USA\\
$^{7}$ CIERA and Department of Physics and Astronomy, Northwestern University, 1800 Sherman Road, Evanston, IL 60201, USA\\
$^{8}$The Adler Planetarium, Chicago, IL 60605, USA\\
$^{9}$ Division of Physics, Mathematics, and Astronomy, California Institute of Technology, Pasadena,USA\\
$^{10}$ DIRAC Institute, Department of Astronomy, University of Washington, 3910 15th Avenue NE, Seattle, WA 98195, USA\\
$^{11}$ Universite Clermont Auvergne, CNRS/IN2P3, LPC, Clermont-Ferrand, France\\
$^{12}$ Caltech Optical Observatories, California Institute of Technology, Pasadena, CA 91125, USA \\
$^{13}$ The Oskar Klein Centre for Cosmoparticle Physics, Department of Astronomy, Stockholm University, AlbaNova, SE-10691 Stockholm, Sweden\\
}

\date{Accepted XXX. Received YYY; in original form ZZZ}

\pubyear{2021}

\begin{document}
\label{firstpage}
\pagerange{\pageref{firstpage}--\pageref{lastpage}}
\maketitle

\begin{abstract}
Type Ia supernovae (SNe~Ia) in the nearby Hubble flow are excellent distance indicators in cosmology. The Zwicky Transient Facility (ZTF) has observed a large sample of supernovae from an untargeted, rolling survey, reaching $20.8, 20.6, 20.3$ mag in $g$ $r$, and $i$-band, respectively. With a FoV of 47 sq.deg, ZTF discovered $>$ 3000 SNe~Ia in a little over 2.5 years. Here, we report on the sample of 761  spectroscopically classified  SNe~Ia from the first year of operations (DR1). The sample has a median redshift $\bar z =$ 0.057, nearly a factor of two higher than the current low-$z$ sample. Our sample has a total of 934 spectra, of which 632 were obtained with the robotic SEDm on Palomar P60.  We assess the potential for precision cosmology for a total of 305 SNe with redshifts from host galaxy spectra.  The sample is already comparable in size to the entire combined literature low-$z$ anchor sample. The median first detection is 13.5 days before maximum light, about 10 days earlier than the median in the literature.  Furthermore, six SNe from our sample are at $D_L < 80$ Mpc, for which host galaxy distances can be obtained in the JWST era, such that we have calibrator and Hubble flow SNe observed with the same instrument. In the entire duration of ZTF-I, we have observed nearly fifty SNe  for which we can obtain calibrator distances, key for percent level distance scale measurements.

\end{abstract}

\begin{keywords}
supernovae:general -- surveys -- distance scale
\end{keywords}



\section{Introduction}
In the two decades since the discovery of the accelerated expansion of the universe \citep{Riess:1998cb,Perlmutter:1998np}, thought to be driven by the as yet poorly understood ``dark energy" component \citep[see][for a review]{2011ARNPS..61..251G}, Type Ia supernovae (SNe~Ia) have been developed to become mature cosmological probes 
\citep{Betoule2014, 2018ApJ...859..101S,Brout18-SMP}. Despite the progress in SN~Ia observations, the nature of the underlying physical mechanism driving acceleration remains an open question, with several potential explanations \citep[see e.g.,][]{Dhawan2017b,2018PhRvD..97d3524L}. Increasing sample sizes of SNe~Ia decreases statistical errors; hence, improving our understanding of dark energy relies on similarly reducing systematic uncertainties. Recent studies find that the systematic and statistical uncertainties on the measurement of the dark energy equation of state, $w$, are approximately equal \citep{2011ApJS..192....1C,2018ApJ...859..101S,Brout18-SMP,2019ApJ...881...19J}.
Apart from measuring dark energy properties, SNe~Ia in the nearby Hubble flow, i.e. redshift range $ z \lesssim 0.1$  (hereafter, also referred to as ``low-$z$") are also critical for measuring the present day expansion rate, i.e. the Hubble constant ($H_0$) precisely. 
Over the last decade, several improvements in the measurements of $H_0$ \citep{2019ApJ...876...85R} have revealed a $>4\sigma$ tension between the local measurement of $H_0$ based on Cepheid calibrated SNe~Ia and the value inferred from the early universe \citep{2018arXiv180706209P}.

This discrepancy could be an indicator of novel cosmological physics, e.g. early dark energy, additional relativistic neutrino species \citep[present a summary of possible explanations]{2020PhRvD.101d3533K}. However, the tension could be caused by poorly understood astrophysical systematics  \citep[e.g. see,][]{scolnic2019,2021arXiv210511461M,2021arXiv210609400M,2021arXiv210615656F}. 
There is significant debate regarding the impact of SN~Ia environments and host galaxy properties on the inferred SN~Ia luminosity, and hence, $H_0$ \citep{rigault2015,Jones_2018,rigault2020}. Conclusively testing the extent of the bias on $H_0$ requires probing the underlying distribution of SN~Ia host galaxy properties,  which is not possible with observing campaigns targeted to specific host galaxy types. Having a sample of SNe~Ia discovered and followed-up by the {\it same}, untargeted survey is, therefore, crucial  to observe the entire range of SN~Ia environmental properties and test for possible biases in inferring cosmological parameters. 


Dark energy inference with SNe~Ia relies on relative distances. The high-$z$ SNe~Ia magnitude redshift relation needs to be ``anchored" at low-$z$. The current low-$z$ anchor sample is compiled heterogeneously from several different photometric systems, each with its own set of systematic uncertainties, which can be correlated in a way that is extremely difficult to predict.  Some of the surveys in the low-$z$ anchor sample targeted pre-selected galaxies \citep[e.g. the Lick Observatory Supernova Survey;][]{2001ASPC..246..121F} and their sample selection and follow-up criteria are heterogeneous and sometimes not well documented. This leads to several systematic uncertainties relating to observational and selection biases. SN~Ia samples at high redshifts ($z \gtrsim 0.1$) e.g. from Sloan Digital Sky Survey \citep[SDSS;][]{kessler09a}, Supernova Legacy Survey \citep[SNLS;][]{2011ApJS..192....1C,Sullivan_2011}, Panoramic Survey Telescope and Rapid Response System  \citep[Pan-STARRS;][]{rest14,2019ApJ...881...19J} and Dark Energy Survey \citep[DES;][]{Brout18-SMP}, however, are homogeneously observed on single photometric systems. Hence, the low-$z$ sample, counter-intuitively, contributes most significantly to the systematics error budget for dark energy inference from SNe~Ia \citep[see also][]{foley2018}. Beyond their use as probes of the expansion history, SNe~Ia have previously been used as tracers of local large scale structure (LSS), e.g. for measuring bulk flows \citep{feindt2013,2016ApJ...827...60M}, the product of the growth rate and amplitude of mass fluctuation, i.e. $f\sigma_8$ \citep{2017JCAP...05..015H}. SNe~Ia are significantly more precise distance indicators compared to methods using galaxy distance, e.g. the Tully-Fisher relation, \citep{kourkchi2020}, and are therefore, an exciting route to infer properties of local LSS. So far, SNe~Ia samples have been very sparse, which is the largest limitation of current datasets for precision  constraints on local LSS with SNe~Ia. 

The advent of modern, wide field surveys has  greatly increased the SN discovery rate  almost by an order of magnitude \citep[e.g., see][]{2014ApJ...788...48S,2018PASP..130f4505T,graham2019,2021ApJ...908..143J}, making it possible to overcome the aforementioned uncertainties.
The Zwicky Transient Facility (ZTF), using the 48-inch telescope at Palomar, is the widest field transient survey in the optical for SN discovery and follow-up. ZTF, in phase-I of operations (hereafter ZTF-I) scanned the entire Northern night sky in the $g$,$r$ filters, reaching $\sim 20.5$ mag/pointing \citep{bellm2019, bellm2019scheduler, graham2019,2020PASP..132c8001D}, with a three-day cadence. To create a definitive data-set of low-$z$ SNe~Ia distances, the main survey was complemented by ZTF partnership surveys, including observations of a large fraction of the sky in the $i$-band. The ZTF SN~Ia survey discovers and provides well-sampled lightcurves, for a large sample of spectroscopically classified SNe~Ia in the $z \lesssim 0.1$ range. This large sample of well-characterised SN~Ia lightcurves from a single, rolling untargeted survey allows for a unique control of the systematics from photometric calibration, observational and astrophysical bias in SN cosmology. This is critical for overcoming the systematics in $H_0$ and dark energy studies.
Moreover, the large statistics and uniform sky coverage, will help overcome existing uncertainties in inferring the growth of structure from SNe~Ia \citep[see e.g.][]{graziani2020}.

In this paper, we present the data-set from the first year of operations and first results from the ZTF SN~Ia survey. The ZTF SNe~Ia sample aims to: (1) anchor current and future high-$z$ SN~Ia sample for dark energy studies, (2) unlock the study of large scale structure in the nearby universe and (3) measure $H_0$ using a unique, self-consistent, calibrator and Hubble-flow sample. With the anticipated deluge of well-observed high-$z$ SNe~Ia from the Vera Rubin Observatory Legacy Survey of Space and time \citep[see][for the science requirements document
]{2018arXiv180901669T}, the large, homogeneously measured, untargeted low-$z$ sample of SNe~Ia from ZTF has the potential to be a definitive data-set for cosmology in the coming decade.

The paper is structured as follows. In section~\ref{sec:data}, we describe the survey strategy, dataset and processing methodology. In section~\ref{sec:result} we describe the analysis and results from the first data release. We present a discussion and our conclusions in section~\ref{sec:discussion}.   

\begin{figure*}
    \centering
    \includegraphics[width=1\textwidth]{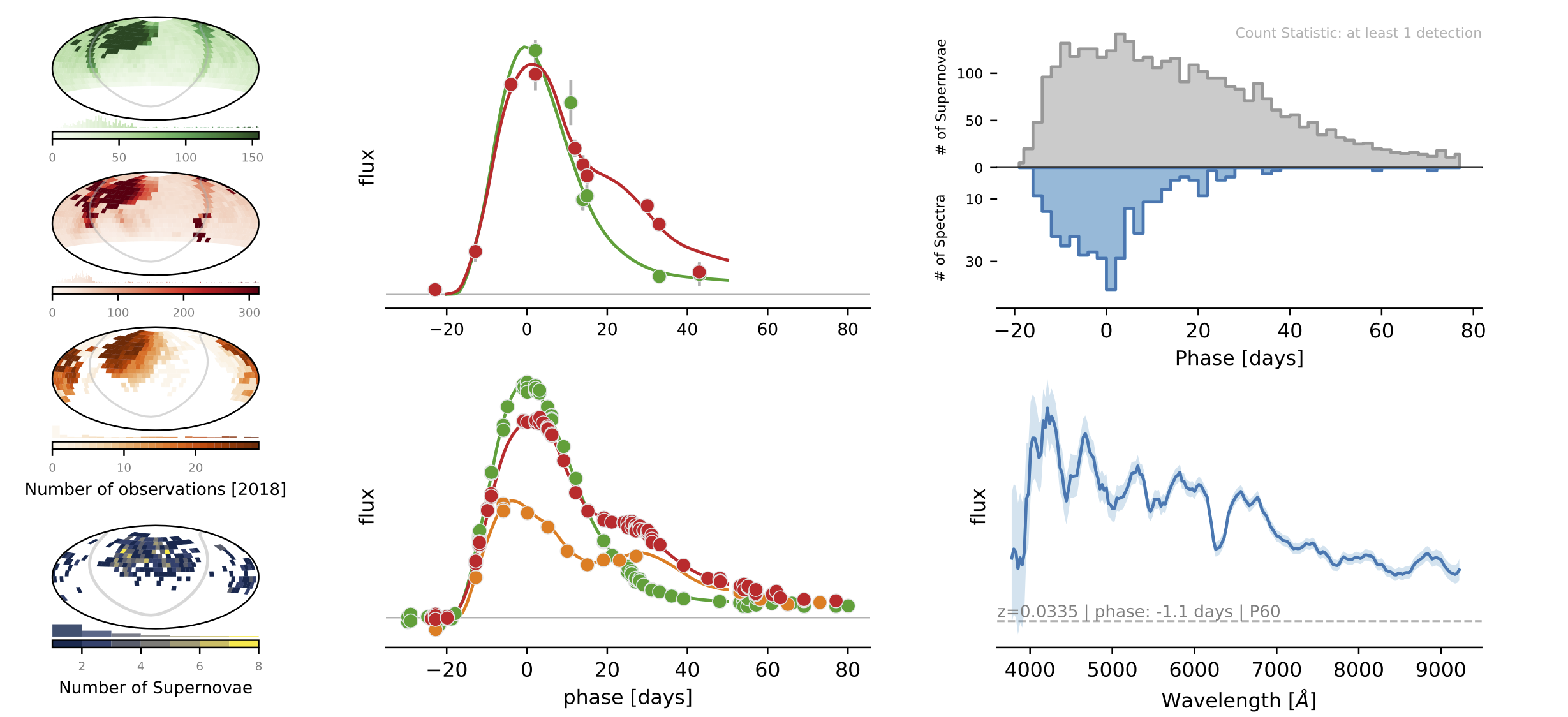}
    \caption{ (Left) The number of observations during the first year of operations in the $g$ (green), $r$ (red), and $i$ (orange) filters. The number of SNe~Ia in each ZTF field are shown on the ZTF field grid in the bottom panel. (Center) An example lightcurve of an SN~Ia at $z \sim 0.07$ and $z \sim 0.03$ to show the typical sampling of the lightcurves in our sample. (Right) A histogram number of SNe with at least one lightcurve point (top) and spectra (bottom) in each phase bin along with an example of the spectrum at $z \sim 0.033$.}
    \label{fig:Yr1_skydist}
\end{figure*}

\section{Survey, Dataset and Methodology}
\label{sec:data}

The survey design and science objectives for ZTF are described in detail in \citet{bellm2019,bellm2019scheduler} and \citet{graham2019}. In this section, we summarise the aspects of the survey relevant for the SN~Ia first data release (hereafter, DR1) sample. This  includes the extragalactic partnership survey, the selection and classification for our sample, association of the host galaxy and the source of the spectroscopic redshift. We also describe the pipeline used for obtaining multi-band photometry for the SNe.

\subsection{Survey Strategy}
ZTF uses a 47 deg$^{2}$ field with a 600 megapixel camera to scan the entire northern visible sky at rates of $\sim 3760$ square degrees per hour  \citep{bellm2019}. This is more than an order of magnitude improvement in survey speed relative to its predecessor survey the Palomar Transient Factory \citep{2009PASP..121.1334R}.  ZTF reaches 5$\sigma$ depth of $\sim$ 20.8 mag in $g$ and $\sim$ 20.6 mag in $r$ band (AB mag) in 30s exposures. The median seeing for the survey is 2.17$^{"}$, 2.01$^{"}$ and 1.82$^{"}$ in the $g,r,i$ bands respectively.

ZTF, in phase-I of observations from March 2018 to November 2020, operated a unique survey strategy with 40$\%$ of its time for public surveys financed by the American's National Science Foundation as part of its Mid-scale Innovations Program (MSIP), 40$\%$ for partnership observations and 20$\%$ for Caltech programs.  In this paper, we focus on the first year of observations, hence, unless mentioned otherwise, we refer to ZTF phase-I as ZTF. The SN~Ia cosmology program derives its dataset from a combination of public and partnership observations. 
During its MSIP time, ZTF observed the entire visible sky from Palomar in $g$ and $r$ filters every three nights with a fiducial exposure time of 30s \citep{fremling2020}. 
The distribution of the SNe across the sky is shown in Figure~\ref{fig:Yr1_skydist}, which also presents a graphical overview of the DR1 survey, including lightcurve and spectral sampling. Crucially for studies pertaining to the measurement of local large scale structure, the SNe are distributed all across the sky \citep[e.g.][]{feindt2013, graziani2020}.

\subsection{The partnership survey}
\label{ssec:partnership}
The MSIP survey is complemented with ZTF partnership surveys. As part of these surveys, high-cadence data was obtained in the $g$ and $r$ bands as well as a wide-area $i$-band survey \citep{bellm2019scheduler,graham2019}.  The high-cadence survey, in the first year of operations, observed a total of $\sim 2500$ deg $^{2}$ with 3 visits per night in the $g$ and $r$ bands, aimed at finding SNe in their infancy. This survey has been critical for early discoveries of SNe~Ia \citep{2019ApJ...886..152Y}, to characterise their rise times \citep{2020ApJ...902...47M} and use the early colours as a test for explosion models and multiple populations \citep{2020ApJ...902...48B}. The survey also provides densely sampled lightcurves in the $g$ and $r$ bands for the sample presented here.

The $i$-band survey involves 30s long exposures with a four-day cadence, over a footprint of 6700 square degrees, approximately one-quarter of the sky area of the MSIP survey, to a median nightly depth of 20.3 mag. The three filter photometry allows us to improve the constraints on the measured colour and is crucial to quantify  colour calibration systematics. Previous studies extending the SN~Ia Hubble diagram in the restframe $i$-band to high-$z$ ($z \sim 0.5$ and higher) find a reduced contribution of systematics errors compared to the optical and argue for independent distances in the $i$-band to the same SN~Ia \citep{2005A&A...437..789N,2009ApJ...704.1036F}. At low-redshifts, observations of SNe~Ia in the $i$-band have also been used to precisely estimate $H_0$ \citep{2018ApJ...869...56B}.
\begin{table*}
\caption{The SNe~Ia in the complete Year 1 sample along with their coordinates, classification, first observation date, instrument from which the spectrum has been obtained, approximate resolution of the instrument and date of first spectroscopic observation. (Table has been truncated to improve the  presentation, the table in its entirety is available online)}
\begin{tabular}{|c|c|c|c|c|c|c|c|c|}
\hline
Name & IAU Name & RA (deg) & Dec (deg) & Classification & Obs Date & Instrument & Resolution & Date of Spectrum \\
\hline
ZTF18aaadqua & SN2018lq & 26.7987 & 18.7986 & SN Ia & 2018-1-14 & P200 & 100-10000 & 20180122 \\
ZTF18aabasml & SN2018tz & 166.4864679 & 37.6251311 & SN Ia & 2018-2-9 & P60 & ~100 & 20180221 \\
ZTF18aabqgnb & SN2018yc & 178.1895182 & 37.8542631 & SN Ia & 2018-2-9 & Lijiang2.4m & ~300 & 20180228 \\
ZTF18aabssme & SN2018aca & 155.7115261 & 14.054786 & SN Ia & 2018-3-5 & P60 & ~100 & 20180307 \\
ZTF18aabstmw & SN2018aaz & 153.4511418 & 38.7630928 & SN Ia 91bg-like & 2018-3-5 & P60 & ~100 & 20180307 \\
\hline
\end{tabular}
\label{tab:summary}
\end{table*}

SNe~Ia show a diverse morphology at late times in the redder wavelengths ($izYJHK$ filters). In the $i$-band, the SNe rebrighten $\sim 2$ weeks after the first peak, showing a characteristic second maximum, which is also an important diagnostic for SN~Ia explosion properties \citep{1996AJ....112.2438H,2010AJ....139..120F}. The double-peaked behaviour over the timescales of a few weeks makes them distinguishable from other SN types. In future photometric SN~Ia surveys, having $i$-band and redder coverage will be crucial to improve photometric classification. The ZTF $gri$ dataset is, therefore, an ideal testbed for such classification algorithms. Moreover, the second maximum in the near infrared (NIR), has also been proposed as an alternate metric for standardisation \citep{2016MNRAS.463.4311S}. Hence, observations in the $i$-band have several advantages for SN cosmology and will be a unique test of systematic errors.

\subsection{Spectroscopy and Sample Selection}
\label{ssec:spec_sample}
In this study, we focus on the sample of spectroscopically classified SNe~Ia discovered in 2018, i.e. the first year of operations for ZTF. To include SNe~Ia that were first detected in 2018 but continued to rise after the end of the year, we conservatively take SNe~Ia that peaked before 1 February 2019. This compilation, hereafter known as the DR1 sample, consists of a total of 761 spectroscopically classified SNe~Ia. These SNe~Ia are on 340 unique ZTF fields corresponding to $\sim 2.2$ SNe per field. Accounting for downtime in the survey operations during the first year, this corresponds to $\sim 2.5$ SNe per field per year. 

A summary of the first year of operations in presented in Figure~\ref{fig:Yr1_skydist}. In the left panels, the ZTF field grid shows the number of $g$ (green), $r$ (red) and $i$ (orange) band observations along with the number of SNe per field.The extragalactic fields with a large number of $g$ and $r$ observations correspond to the high-cadence survey described in section~\ref{ssec:partnership}. We show an example lightcurve of an SN~Ia with only MSIP and one with MSIP and partnership data to illustrate the typical sampling in the three filters. We show a histogram distribution of the number of SNe with at least one observation for each phase with a 1 day bin size. We find a significant fraction of SNe~Ia have at least one observations before -7 days.
\begin{table}
\caption{SN names, heliocentric frame redshift and  coordinates (in degrees) for the SNe~Ia in our sample. The table has been truncated for formatting reasons. Full table is available online. 
}
\begin{tabular}{|c|c|c|c|}
\hline 
SN Name  & $z_{\rm helio}$ & RA & Dec  \\
& & (Deg) & (Deg)  \\
\hline
ZTF18aabdgik & 0.021604 & 175.5987 & 10.2641 \\
ZTF18aabstmw & 0.02308648 & 153.4511 & 38.7631 \\
ZTF18aabsyqp & 0.07667208 & 162.8185 & 22.4776 \\
ZTF18aabtaor & 0.07983481 & 172.0952 & 44.7490 \\
ZTF18aabxrjp & 0.07917714 & 185.5432 & 26.9946 \\

\hline
\end{tabular}
\label{tab:coords_z}
\end{table}

\begin{figure*}
    \centering
    \includegraphics[width=.9\textwidth]{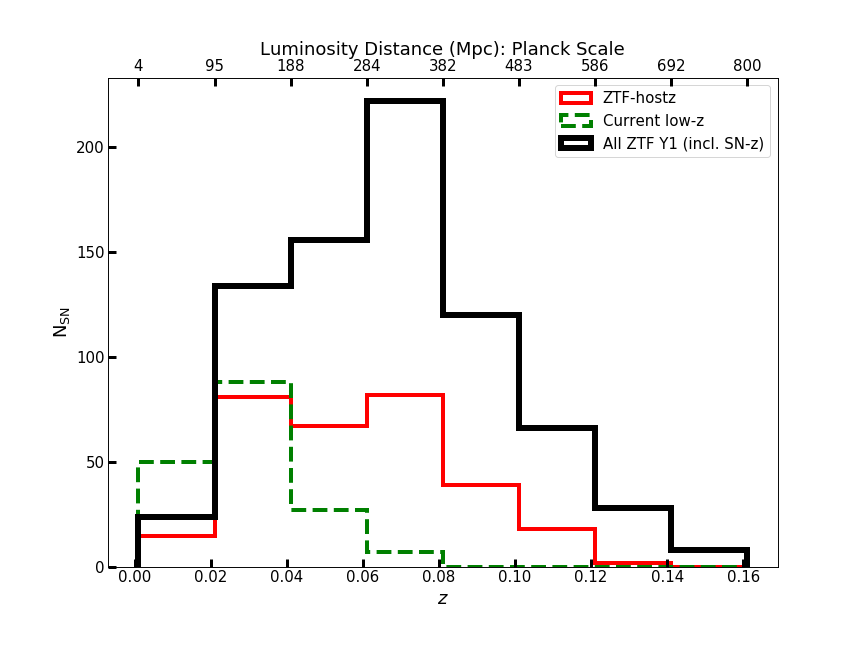}
    \caption{Histogram distribution of the redshifts from the DR1 sample with host galaxy redshifts (red) and the total sample with redshifts from any source, including fitting the SNID template to the SN spectrum (black) compared to the current low-$z$ sample in the literature (green). The redshift distribution of our DR1 sample has a median $z = 0.057$ that is $\sim$ twice higher than the median from the literature sample, hence, making this sample significantly less sensitive to the uncertainties from peculiar velocity corrections.}
    
    \label{fig:redshift_dist}
\end{figure*}

A large fraction of these SNe~Ia were classified using the SEDmachine \citep{2012SPIE.8446E..86B, 2018PASP..130c5003B, 2019A&A...627A.115R}, a low-resolution (R $\sim 100$) integral field unit (IFU) spectrograph on the robotic Palomar 60$^{"}$ telescope \citep{2006PASP..118.1396C} as part of the ZTF Bright Transient Survey (BTS), with the goal to spectroscopically classify and publicly report every extragalactic transient in the Northern sky with $r < 18.5$ mag discovered by the ZTF public surveys. The SEDm is capable of classifying $> 10$ SNe (of any type) per night in the 18.5 - 19 magnitude range. Details of the BTS program can be found in \citet{fremling2020,2020ApJ...904...35P}. The data are reduced uniformly and consistently as part of the automated \texttt{PYSEDM} pipeline \citep{2019A&A...627A.115R}.

For the complete sample of 761 objects, we have obtained a total of 934 spectra, hosted on the GROWTH marshal \citep{2019PASP..131c8003K}. Out of the total, 632 spectra are obtained with the SED machine, which corresponds to $\sim 68\%$ of the total sample of spectra. We present a distribution of the phase at which the spectra were obtained in Figure~\ref{fig:Yr1_skydist} (top right). A example of a high S/N spectrum for an SN at $z = 0.0335$ in shown in the bottom right panel of Figure~\ref{fig:Yr1_skydist}. 

\subsection{Host galaxy redshifts}
\label{ssec:host-z}
In our study, we want to infer the lightcurve parameters for the SNe~Ia in the DR1 sample and quantify the scatter in the Hubble residuals. This requires a robust, spectroscopic determination of the redshift to the SN host galaxy. We compile the SN~Ia host galaxy redshifts here. A large fraction of these host galaxy redshifts are obtained from the Sloan Digital Sky Survey (SDSS) sixteenth data release \citep{2020ApJS..249....3A}. For SNe~Ia that do not have a redshift for the host galaxy in SDSS, we obtain the redshift from the NASA Extragalactic Database (NED)\footnote{\url{https://ned.ipac.caltech.edu/}}. In addition to  host galaxy redshift constraints, we restrict the sample to SNe with sufficient data around and before maximum light from the alert pipeline. This is required to have a robust initial estimate of the time of maximum for creating custom reference and difference images to perform forced photometry (see section~\ref{ssec:phot_pipe} for details). The final sample with these constraints has 305 SNe~Ia. The coordinates and redshifts for SNe~Ia in this sample, are presented in Table~\ref{tab:coords_z}. Hereafter, we refer to this as the host-$z$ sample.
 
We emphasise that while this criterion selects less than half ($\sim 40\%$) the total sample of SNe~Ia \citep[consistent with the findings of BTS]{fremling2020}, host galaxy redshifts are not time critical. These can be obtained after the completion of the survey. Several surveys using multi-object spectrographs e.g. the Dark Energy Spectroscopic Instrument \citep[DESI;][]{2016arXiv161100036D} and/or the 4MOST consortium extragalactic surveys  \citep{2019Msngr.175....3D,2019Msngr.175...58S} can obtain spectra for galaxies that hosted ZTF observed SNe~Ia, from which redshifts can be derived. Despite the incompleteness of available host redshifts, we note that even the host-$z$ sample is larger than the entire, combined low-$z$ anchor in current SN cosmology studies \citep[e.g.][]{2018ApJ...859..101S}.

Our sample presented here has a median redshift of 0.057 which is nearly twice that of the median for the sample in the literature. Hence, the inferred cosmological parameters from this sample will be less prone to systematic errors from peculiar velocity corrections \citep[see, e.g.][]{2020arXiv201005765H}. The resulting redshift distribution for our sample is shown in Figure~\ref{fig:redshift_dist}. For the complete sample as well as the host-$z$ only sample, we can see that median redshift is $\sim$ twice the median of the current literature low-$z$ sample (with All ZTF Y1 having a higher median $z$ of 0.069).

\subsection{Photometry pipeline}
\label{ssec:phot_pipe}
In this section, we describe the pipeline used for generating the photometry for our sample of SNe~Ia. Since we want to measure lightcurve model parameters and Hubble residuals, we focus on the host-$z$ sample. While this is a custom-based pipeline for creating reference images, image subtraction and source photometry, it uses certain aspects for the ZTF data products made available by the Image Processing and Analysis Center (IPAC) at Caltech, hence, we summarise those aspects here. 
The ZTF data processing, products, and archive are described in detail in \citet{2019PASP..131a8003M}. IPAC provides alert packets with photometry as well as a forced photometry service at a predetermined locations. We note that for a substantial fraction of the DR1 sample, building of reference images began during the survey operations and hence, overlapped in time with the discovery of astrophysical transients. Therefore, the alert and IPAC forced photometry for several SNe~Ia had contamination from the live transient in the reference image. This  required a reprocessing of the data to make customised references for the regions of the sky where the SNe~Ia were discovered.  

We use a state-of-the-art image processing pipeline, written using well-tested and widely used astronomical image processing tools for processing data from the ZTF Uniform Depth Survey (ZUDS) \footnote{\url{https://github.com/zuds-survey/zuds-pipeline}}. While the pipeline was designed for discovering slowly evolving faint transients on deep coadds of science frames, its versatile functionality makes it ideal for creating image subtractions and computing lightcurves using single exposures. Details of the individual components of the pipeline are presented in Appendix~\ref{sec:pipe_desc}. We summarise the pipeline features in this section.

Using the ZUDS pipeline, we create references for each ZTF field and CCD in which an SN~Ia in our sample has been observed \citep[see section 3 of ][for details about the ZTF CCDs]{2019PASP..131a8003M}. 
They are generated using the \texttt{SWARP} \citep{2002ASPC..281..228B} software for coaddition. The associated reference mask is created with the mask images provided for each exposure by IPAC, using a bitwise logical AND (see Appendix~\ref{sec:pipe_desc}). 
When sufficient number of exposures are available, reference images are created by coadding epochs from before the SN explodes (which we determine as epochs 30 or more days before the time of maximum inferred from photometry provided by IPAC). For making the reference image we only use images that were obtained in seeing conditions between 1.7$^{"}$and 3$^{"}$ and have a deeper magnitude limit than 19.2 mag. If there are insufficient number of exposures ($< 20$) for which the nightly conditions do not meet these criteria, the pre-SN exposures were combined with post-explosion images from $> 400$ days after the time of maximum inferred from the alert photometry.

\begin{figure}
    \centering
    \includegraphics[width=.48\textwidth]{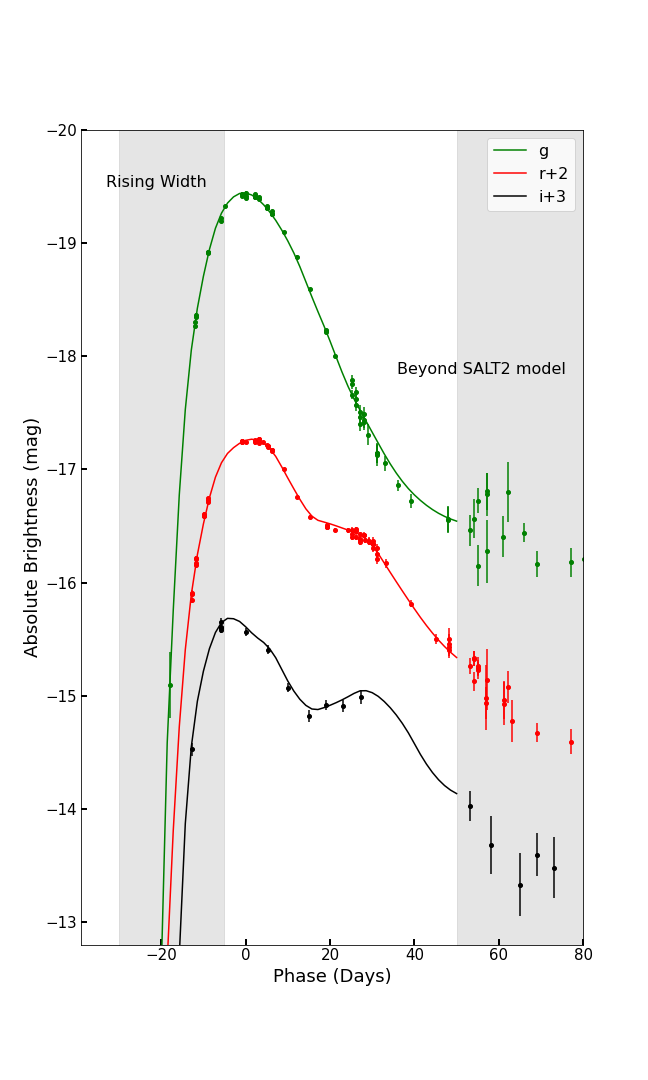}
    \vspace{-6mm}
    \caption{An example of fitting the SALT2 model to ZTF18aaumeys, as implemented in \texttt{sncosmo}. The SN is at redshift 0.033, well into the Hubble flow and the lightcurve is sampled extremely well on the rising part, which also allows us to explore beyond the conventional standardisation parameters. Data for this SN extends beyond +50 days, the validity limit of the SALT2 model. Hence, we can use SNe in our sample to retrain the SALT2 model and extend to later phases.  We note here that the SALT2 model has only been fitted to the $g$ and $r$ band data, whereas the for $i$-band only the predicted model lightcurve is overplotted on the data. }
    \label{fig:example_saltfit}
\end{figure}
We generate difference images using the \texttt{HOTPANTS} image subtraction software \citep{2015ascl.soft04004B}.  
The subtraction is normalised to the science image and convolved to the template image. Details of the convolution kernel, half width substamp, and the setup for the \texttt{HOTPANTS} subtractions are described in Section~\ref{sec:pipe_desc}.

We use the \texttt{PhotUtils} astropy package \citep{Bradley_2019_2533376} for performing aperture photometry at the SN position, determined from the alert photometry provided by IPAC.
For each epoch, the zero point is derived from a combination of the nightly zero point and aperture correction provided by IPAC \citep[see][for details]{2019PASP..131a8003M}.

\begin{figure*}
    \centering
    \includegraphics[width=.9\textwidth, trim=20 50 20 0]{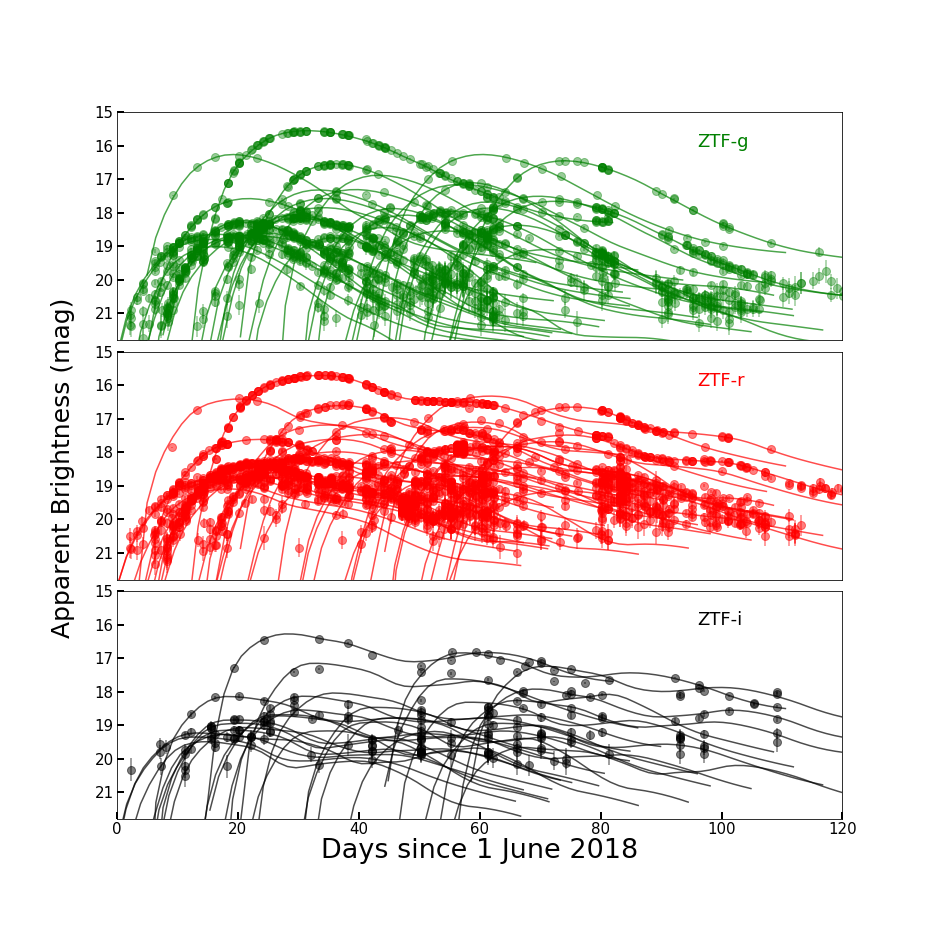}
    \caption{Lightcurves for the DR1 host-$z$ sample in the $g$,$r$,$i$ filters for SNe~Ia that peak within 2 months of 15 June 2018 (shown as days from 1 June 2018). The SALT2 model fits to  only the $g$ and $r$ band data are overplotted as solid lines.}
    \label{fig:lightcurve_completeplot}
\end{figure*}

\section{Analysis and Results}
\label{sec:result}

We analyse the host-$z$ sample in this section, describe the spectroscopic properties and the sample demographics, in comparison with the literature sample. We also present the sample of SNe~Ia that are at feasible distances to build a calibrator sample for local $H_0$ measurements.

\subsection{Lightcurve fitting}
\label{ssec:lc_fit}
To be used for cosmology, SNe~Ia need to be standardised, using relations between their luminosity and the light curve shape and colour, for measuring accurate distances  \citep[see][for a review of how SNe~Ia are used in cosmology]{2018SSRv..214...57L}. Here we describe the procedure for fitting the lightcurves  generated using the pipeline described in section~\ref{ssec:phot_pipe}, to derive the peak apparent brightness, lightcurve width and colour. 

There are several algorithms in the literature to estimate distances from SN lightcurves, e.g. MLCS \citep{2007ApJ...659..122J}, SiFTO \citep{2008ApJ...681..482C}, BayeSN \citep{2009ApJ...704..629M,mandel2011,2020arXiv200807538M}, SNooPy \citep{burns2011}, BaSALT \citep{2014ApJ...780...37S}. Each lightcurve fitting method makes a correction to the SN luminosity for the lightcurve shape and colour. The colour correction, however, can be applied in two principly different ways. The first uses an empirical correlation between the observed colour and the uncorrected Hubble residuals, whereas the second assumes that the observed colour is a combination of intrinsic SN colour, photometric errors and reddening due to dust, e.g. in the SN~Ia host galaxy. The former is more model independent, whereas the latter is more physically motivated. In-depth comparisons of different lightcurve fitters in the literature \citep[e.g.][]{kessler09a}, find that in general they agree reasonably well given the same input assumptions. 

Currently, the most widely used lightcurve fitting algorithm is the Spectral Adaptive Lightcurve Template - 2 \citep[SALT2;][]{guy2007}, based on the SALT method \citep{2005A&A...443..781G} and we use this in our analysis. The SALT2 model treats the colour entirely empirically and is used to find a global colour-luminosity relation. We use the most updated, published version of SALT2 \citep[SALT2.4, see][]{guy2010} as implemented in \texttt{sncosmo} v2.1.0 \footnote{\url{https://sncosmo.readthedocs.io/en/v2.1.x/}} \citep{2016ascl.soft11017B}. In the fitting procedure, we correct the SN fluxes for extinction due to dust in the Milky Way (MW). We use extinction values for the SN coordinates (presented in Table~\ref{tab:coords_z}) derived in \citet{2011ApJ...737..103S}. These are updated galactic extinction maps that find that the extinction values from \citet{1998ApJ...500..525S} were overestimated. We use the widely applied galactic reddening law, proposed in \citep{1989ApJ...345..245C}, known as the ``CCM" law to correct for MW extinction. For the extinction corrections, we use the canonical value for the total-to-selective absorption, $R_V = 3.1$.
The SALT2 model assumes a parametrisation from \citet{tripp1998}
\begin{equation}
    \mu_{\rm B} = m_{\rm B} + \alpha x_1 - \beta c - M_{\rm B}
    \label{eq:tripp}
\end{equation}

\begin{figure*}
    \centering
     \includegraphics[width=.48\textwidth]{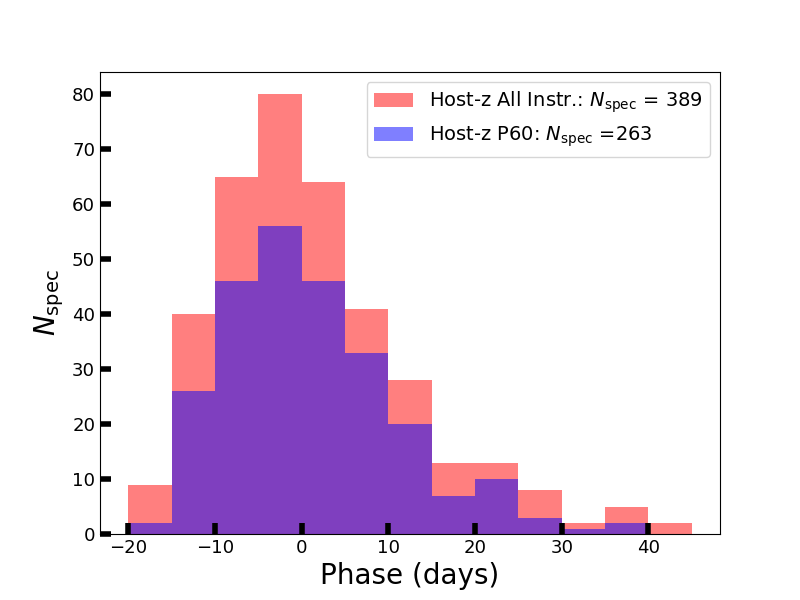}
    \includegraphics[width=.48\textwidth]{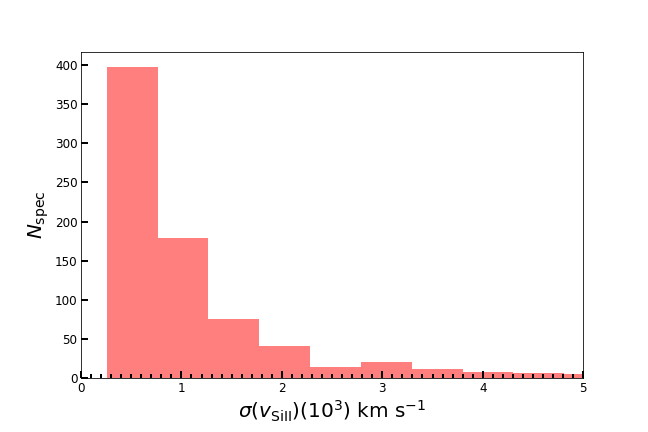}
    \caption{(Left): Phase distribution of the spectra in the host-$z$ sample from all instruments (red) and from SEDm alone (blue). Similar to the complete DR1 sample, $68\%$ of all spectra are obtained using SEDm. (Right) The histogram distribution of the errors on the Si II 6355\AA\, line velocity (in km\,s$^{-1}$) from only the SEDm spectra for the SNe in our sample, inferred using the \texttt{spextractor} software. We find a median error of $\sim 700$ km s$^{-1}$.}
    \label{fig:spectra_Y1sample}
\end{figure*}
where $\mu_{\rm B}$ is the distance modulus inferred from restframe $B$-band parameters, determined for each SN given the observed $B$-band brightness, $m_{\rm B}$, light-curve shape, $x_1$, and an observed colour, $c$. $\alpha$, $\beta$ and $M_{\rm B}$ are nuisance parameters, namely, the slope of the width-luminosity and colour luminosity relations and the absolute magnitude of the SNe~Ia in the $B$-band for a fiducial SN~Ia with $x_1 = c = 0$. In our analyses, we do not use any independent, absolute calibration for the SN~Ia luminosity, hence, the $M_B$ term is effectively, only the intercept of the magnitude-redshift relation, $a_B$. All nuisance parameters are globally fit for the sample. We note that since the SALT2.4 model is not appropriate for wavelengths redder than 7000 \AA\, we only fit the $g$ and $r$ band lightcurves for the host-$z$ sample.

We fit the SALT2 model to each SN~Ia in two steps. Firstly, we fit without the SALT2 model covariance to obtain an initial guess for the time of maximum, $t_0$. We select the data in the phase range -20 to +50 days relative to the initial guess $t_0$, corresponding to the phase range in which the SALT2 model is defined. For the fit, we use an error floor of 2$\%$ of the flux, corresponding to the the typical photometric calibration accuracy with respect to Pan-STARRS1 reported in \citet{2019PASP..131a8003M}.  We note that this is appropriate since we use the zeropoint report by \texttt{IPAC} in \citet{2019PASP..131a8003M} for fitting the fluxes computed from the photometry.

We then fit the data only in this phase range along with the model covariance to get the final fit parameters. The resulting parameters are reported in Table~\ref{tab:salt_params}.
As we can see in Figure~\ref{fig:example_saltfit}, ZTF18aaumeys, shown as an illustrative example SN, has a significant number of detections on the rising part of the lightcurve. This can be used to improve the standardisation relative to just the standard stretch and colour corrections (as in eq.~\ref{eq:tripp}), and ultimately reduce the intrinsic dispersion in the sample. It has been proposed in the literature that separating the lightcurve width parameter $x_1$, into an $x_1$ from the rising and falling parts of the lightcurve, $x_1^r$ (labeled as the rising width in Figure~\ref{fig:example_saltfit}) and $x_1^f$, can improve the lightcurve fit. Since it has been shown that for a given decline rate, the rise times vary significantly \citep{2011MNRAS.416.2607G,2015MNRAS.446.3895F,2020ApJ...902...47M}, we expect to gain in the precision of the distance measurement by using information in rising part of the lightcurve. It has been shown that this method of two lightcurve widths instead of one yields a smaller scatter than if only using a single lightcurve width \citep{2019ApJ...871..219H}.
ZTF18aaumeys is also an example SNe~Ia that has data extending beyond the range of phases for which the SALT2 model is valid, which makes it interesting to use this data for further extending the SALT2 model beyond the phase range of +50\,days. The dataset presented here allows us to test these novel standardisation procedures in the future.

\begin{table*}
\caption{Output SALT2 fit parameters for SNe in our sample along with the CMB frame redshift ($z_{\rm CMB}$; see text for the CMB frame conversion). We report the $m_B$, $x_1$ and $c$ parameters used to compute the Hubble residuals (Full table available online; {\bf fit outputs including covariances between the fit parameters are provided in the github repo})}
\begin{tabular}{|c|c|c|c|c|}
\hline
ZTF Name & $z_{\rm CMB}$ & $m_B$ &  $x_1$  & $c$  \\
&& (mag) && \\
\hline
ZTF18aabdgik & 0.022775296 & 15.516 ($
pm$ 0.047) & 0.096 ($\pm$ 0.218) & 0.034 ($
\pm$ 0.038) \\
ZTF18aabstmw & 0.023919309 & 18.005 ($\pm$ 0.066) & -1.302 ($\pm$ 0.148) & 0.433 ($\pm$ 0.048) \\
ZTF18aabsyqp & 0.077744924 & 18.579 ($\pm$ 0.121) & 0.556 ($\pm$ 0.221) & 0.193 ($\pm$ 0.091) \\
ZTF18aabtaor & 0.080599043 & 18.397 ($\pm$ 0.075) & 0.659 ($\pm$ 0.451) & -0.045 ($\pm$ 0.057) \\
ZTF18aabxrjp & 0.080152526 & 18.479 ($\pm$ 0.070) & -0.732 ($\pm$ 0.433) & 0.022 ($\pm$ 0.053) \\
\hline
\end{tabular}
\label{tab:salt_params}
\end{table*}

\subsection{The $gri$-sample}

For the host-$z$ sample, we have a total of 170 SNe~Ia with at least one observation in the $i$-band between -15 and +15 days relative to the $T_{\max}$ inferred from the SALT2 fits to the $g$ and $r$ band data. However, to get robust detections, we need deep references to create good difference images. Therefore, we further require at least 15 frames in the $i$-band either the phase range before -30 days or after +400 days to make the reference image. 
Hence, for the $i$-band, we further impose the cut of the number of frames to create a reference image. We note that for our host-$z$ sample, out of the 305 SNe, 122 have at least 1 detection in the $i$-band. This corresponds to $\sim 40\%$ of the SNe in the sample, which is consistent with the predictions from \citet{2019JCAP...10..005F}. We emphasize that this is a lower limit, with the number expected to increase with an increase in the number of observations after the SN has faded, for making a reference image. 

The lightcurve fits for the SNe that peaked within 60 days since 15 June 2018 are shown in Figure~\ref{fig:lightcurve_completeplot}. We note that the fit is only to the $g$ and $r$-band data and the solid black lines in the bottom panel are the SALT2 model predictions in the $i$-band overplotted on the data. The sample parameters are discussed in Section~\ref{ssec:sample_dem} along with further selection cuts in Table~\ref{tab:sample_select}.

\subsection{Spectroscopic properties}
The sample presented here is comprised entirely of spectroscopically classified SNe~Ia. This is critical to remove contamination from core-collapse SNe and other non-SN~Ia transients. Several studies in the literature have demonstrated that SNe~Ia can be divided into subclasses based on spectral line velocities near maximum light, with potential implications for distance measurements  \citep{2009ApJ...699L.139W,2011ApJ...729...55F,2013ApJ...773...53F,2020MNRAS.493.5713S,2021arXiv210206524D}. Therefore, having spectra is important for improving cosmological inference with SNe~Ia. The phase distribution for the spectra of the SNe~Ia in the host-$z$ sample is shown in Figure~\ref{fig:spectra_Y1sample}. We find no significant difference in the phase distribution of the spectra from all instruments and the SEDm alone. 50$\%$ of all spectra (and 50$\%$ of all SEDm spectra) are obtained before maximum light, while 23$\%$ (20$\%$ for SEDm) are obtained before -7\,d. Here, measure the accuracy of the line velocities from spectra obtained with the SEDm. 

Since we aim to measure the accuracy of line velocities, we require an independent determination of the redshift and hence, we use the host-$z$ sample of 305 SNe~Ia, which is a lower limit since the host redshift are not time critical (see section~\ref{ssec:host-z}). 
In this analysis, we only focus on the ubiquitous Si II 6355 \AA\, feature, since the velocity of this feature has been most widely proposed in the literature as a metric for subclassifying SNe and improving distances. We reiterate that a large fraction (68$\%$) of the spectra for our sample are obtained with a single instrument, i.e. SEDmachine, extracted uniformly with the \texttt{PySEDM} pipeline \citep{2019A&A...627A.115R}.

For our line velocity measurements we use a fully automated and public code for spectral fitting, \texttt{spextractor}\citep{Papadogiannakis1300923}~\footnote{\url{https://github.com/astrobarn/spextractor}}. The code uses a non-parametric, gaussian process (GP) regression \citep{Rasmussen:2005:GPM:1162254} to get the minima for the individual features. It is based on the publically available \texttt{python} package GPy \citep{gpy2014} and uses a Matern 3/2 kernel for smoothing the spectra.

The median error in the inferred $v_{\rm Si}$ is $\sim 700$ km s$^{-1}$. We emphasize that this method has fewer model assumptions than methods used in the literature. Hence, we expect the median errors to be higher compared to e.g. a fit assuming multiple gaussians describe the line profile. However, even with such precision on the velocity measurements, we can clearly distinguish high- and low-velocity subtypes to test improvements in the distance measurements with our sample.
A detailed analysis of the method and the spectral sample, as well as relations with lightcurve and host galaxy parameters will be presented in a followup study (Johansson et al. in prep). 

\subsection{Sample cuts and demographics}
\label{ssec:sample_dem}
In this section, we describe quality cuts on the host-$z$ sample. We  perform cuts based on the SN~Ia spectral subtyping, redshift range and lightcurves parameters to get the final sample with which to compute the Hubble residuals. We then compare the lightcurve fit parameters and the Hubble residuals to the current low-$z$ from the literature.

The first selection cut is to remove peculiar SNe from the final analyses. These include SNe~Ia that are spectroscopically similar to the class of SNe~Iax \citep{2006AJ....132..189J,2013ApJ...767...57F} the eponymous subluminous and fast-declining  SN~1991bg \citep{1992AJ....104.1543F,leibundgut1993} or Ia-CSM, similar to the prototype SN~2002ic \citep{2003Natur.424..651H,2004ApJ...616..339W}. This is because such peculiar SNe  might not follow the width-luminosity relation \citep{phillips1993} or they may not be adequately described by the SALT2 model. 
However, we do not remove SNe~Ia in the subclass similar to SN~1991T.
This is for two reasons. Their peculiarities are shown to be almost exclusively spectroscopic, and they are shown to obey the width-luminosity relation \citep[see][for a discussion on the extremes of thermonuclear supernovae]{2017hsn..book..317T}. 
Furthermore, we note that 91bg-like objects might not be inherently incalibratable,  since there are  fitting algorithms other than SALT2, e.g. MLCS2k2 \citep{2007ApJ...659..122J} with which it is possible that can estimate distances for this subclass of SNe~Ia, however, for our current analyses we remove them from the cosmological sample, but possibly might include them in the future. 
\begin{figure}
    \centering
    \includegraphics[width=.48\textwidth, height=14cm, trim = 0 30 0 30 ]{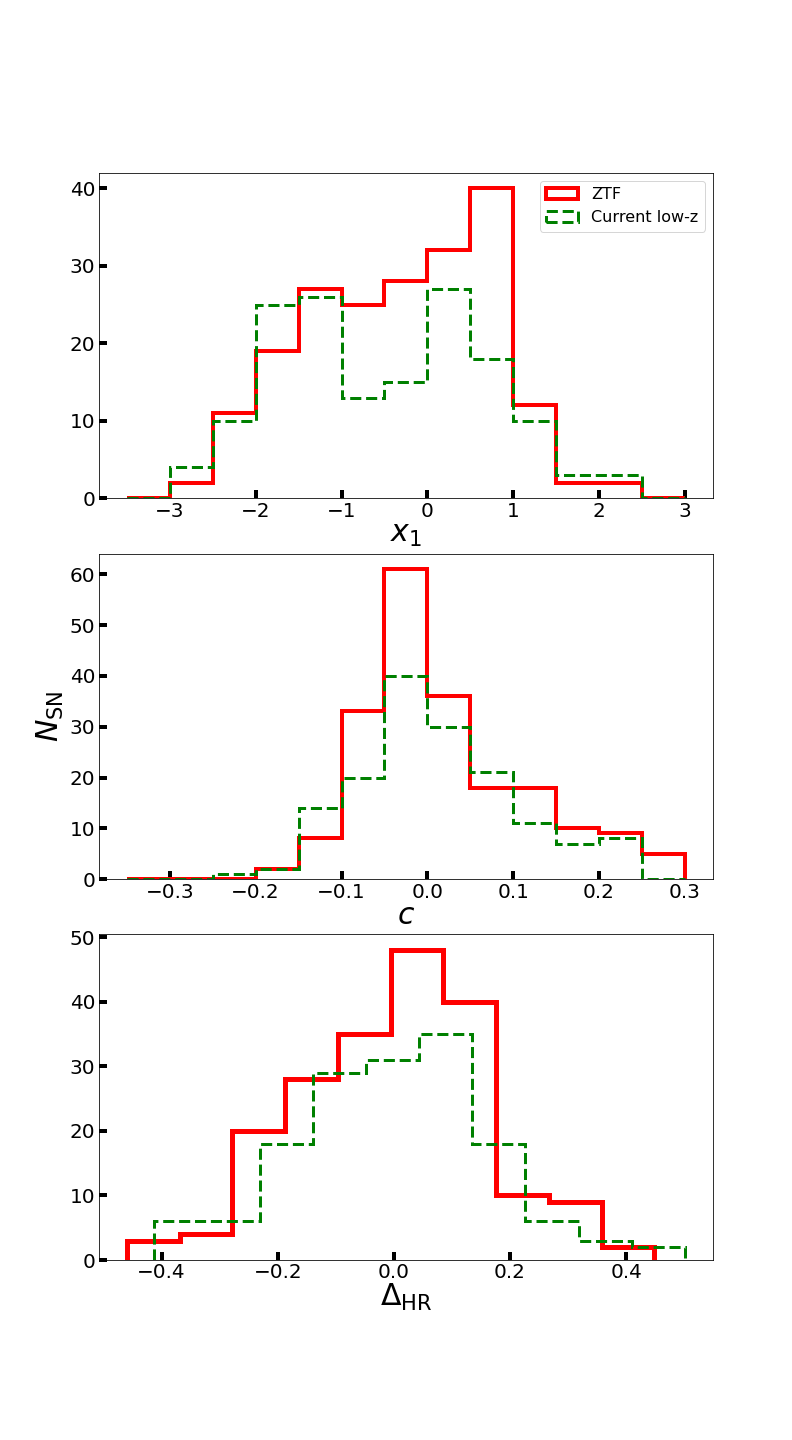}
    \vspace{-6mm}
    \caption{(Top) SALT2 lightcurve width ($x_1$), (middle) colour ($c$) parameters and (bottom) the Hubble residuals computed after fitting the nuisance parameters for the ZTF DR1 sample (red, solid) compared to the low-z sample used for cosmological studies in the literature (green, dashed). We find that the $x_1$-distributions for the ZTF and literature sample appear to be drawn from different parent populations, however, the $c$-distribution for the ZTF sample, within the $|c| > 0.3$ does not. We note, however, that the sample has a larger fraction of $c> 0.3$ SNe~Ia compared to the literature (see Table~\ref{tab:sample_select}) The $\sigma_{\rm rms}$ of the Hubble residuals is 0.17 mag, which is very similar to the $\sigma_{\rm rms}$ of the literature sample.}
    \label{fig:salt_param_dist}
\end{figure}

We apply cuts based on the redshift of the host galaxy. Since this sample is uniquely discovered, followed-up and characterised with a single telescope and instrument, we do not have an a priori truncation on the redshift range. We apply a lower limit on the redshift distribution of $z \geq 0.015$, to minimize errors from peculiar velocity corrections, as done in previous studies in the literature \citep[for e.g. see][]{foley2018}.
The ZTF BTS reports a completeness of 97$\%$ at $m < 18$ mag and 93$\%$ $m < 18.5$ mag \citep{2020ApJ...904...35P}. The ZTF BTS is $\gtrsim 70\%$ complete till $m < 19$ mag, corresponding to a $z \lesssim 0.1$ threshold, hence, we apply an upper limit on the redshift of $z \leq 0.1$.  
\begin{table}
\caption{Selection criteria for the DR1 sample. The final sample, after quality cuts is used to compute the Hubble residuals. The strongest cut is for an available host galaxy redshift, which are not time-critical.}
\resizebox{.47\textwidth}{!}{\begin{tabular}{|l|c|c|r}
\hline 
Criterion & $\#$ of SNe & Cumulative $\#$ of SNe & Remaining SNe \\
& not passing & not passing & \\
\hline
{\bf All DR1}  & $\ldots$ & $\ldots$ & {\bf 761}\\
Host-$z$ & 456 & 456 & 305 \\
Spectroscopically peculiar, & 11 & 467 & 294 \\
$z >= 0.015$, & 5 & 472 & 289 \\
$z <= 0.1$, & 23 & 495 & 266 \\
$> 3$ points between -10 and +10 days & 26 & 521 & 240 \\
$\sigma(x_1) < 1$, $\sigma(t_0) < 1$ & 11 & 532 & 229 \\
$-3 < x_1 < 3$ & 2 & 534 & 227 \\
$-0.3 < c < 0.3$ & 25 & 559 & 202 \\
Chauvenet's criterion ($> 4 \sigma$ outlier) & 2 & 561 & 200 \\

\hline
\end{tabular}}
\label{tab:sample_select}
\end{table}

In our analysis we want to robustly determine the lightcurve fit parameters, hence, we restrict the sample to SNe with at least 3 points in the lightcurve within 10 days of maximum light. Furthermore we remove poor SALT fits, defined as having $\sigma(x_1) > 1$ or $\sigma(t_0) > 1$ day. 
 For the final sample, we also remove SNe with extremely slow or fast declining lightcurves, i.e. $|x_1| > 3$  and SNe with extremely blue or red colours, i.e. $|c| > 0.3$. We note that these SNe are of interest to SN~Ia explosion physics, environmental properties and reddening due to dust in the host galaxy \citep[e.g.][]{2018MNRAS.479.3663B}, but are removed when computing the Hubble residuals. Finally, we apply Chauvenet's criterion on the sample. The selection cuts are summarised below
\begin{itemize}
    \item Spectroscopically classified as a normal SN~Ia, i.e. not spectroscopically similar to SNe~Iax
    \citep{2006AJ....132..189J,2013ApJ...767...57F}, SN~1991bg \citep{1992AJ....104.1543F,leibundgut1993} or Ia-CSM \citep{2003Natur.424..651H,2004ApJ...616..339W}.
    \item $0.015 \leq z \leq 0.1$
    \item At least 3 points between -10 and +10 days 
    \item The uncertainty on $x_1$ is $< 1$
    \item $-3 < x_1 < 3$
    \item $-0.3 < c < 0.3$
    \item Chauvenet's criterion to exclude systematic outliers. 
\end{itemize}
Here, we describe the objects in our sample removed with each of the above selection cuts. 
Our sample includes a total of 11 spectroscopically peculiar SNe~Ia, 9 of which belong to the subclass of 91bg-likes, 1 Type Iax SN and 1 SN~Ia-CSM. For the 91bg-likes which have sufficient coverage between -10 and +10 days, we derive the SALT2 lightcurve parameters. For 5 of the 7 SNe~Ia, the SALT2 $c > 0.3$, consistent with the observation in the literature that 91bg-likes are redder than normal SNe~Ia.

We remove 27 SNe which don't have enough observations near maximum light and 9 SNe which do not have an accurately measured $x_1$. Only 2 SNe are outside the range of $x_1$ values for cosmologically viable SNe~Ia. We find 25 SNe with $|c| > 0.3$, 23 of which are reddened with a $c > 0.3$. We find more reddened SNe in our sample than in other low-$z$, cosmological samples in the literature. Additionally, even within the cosmological sample, there are more SNe~Ia in the range $0.25 < c < 0.3$ compared to the existing low-$z$. The larger number of high-$c$ SNe could be due to the fact that our sample is derived from an untargeted survey, where we don't a priori reject reddened SNe when building a lightcurve, however, we will explore this in detail in future works.
We finally apply Chauvenet's criterion and find only 2 SNe removed with this cut, ZTF18ablqkud and ZTF18aarcypa. They have a $\delta \mu$ of 0.76 and 0.77 mag respectively, making them both significantly fainter than the other SNe in the sample. While faint, ZTF18ablqkud does not show an spectroscopic features similar to the class of SN~1991bg-like SNe and has a broad $r$-band lightcurve, which is consistent with normal SN~Ia. It has an inferred  $x_1 = 0.88 \pm 0.22$ and $c = 0.129 \pm 0.034$ which are also consistent with the values for normal SNe~Ia. ZTF18aarcypa similarly does not show spectroscopic features similar to SN~1991bg, and it shows a shoulder in the $r$-band, however, it has a low $x_1 = -2.89 \pm 0.29$, which is consistent with the class of transitional fast decliners \citep{2015A&A...578A...9H}. A summary of the number of objects removed after each cut is presented in Table~\ref{tab:sample_select}.

We compare the $x_1$ and $c$ distribution for our sample (top and middle panel Figure~\ref{fig:salt_param_dist}) with the low-$z$ anchor sample for dark energy studies in the literature \citep{2018ApJ...859..101S}. Using a Kolmogorov-Smirnov test we find a $p-$ value of 0.03 and 0.6 for the $x_1$ and $c$ distribution. For the $x_1$ parameter, we can tentatively reject, the hypothesis that both distributions are drawn from the same parent population. We will investigate this in detail in future works.  We cannot  reject the null hypothesis that both $c$ distributions come from the same parent population, however, we find a larger fraction of $c > 0.3$ SNe~Ia in our sample compared to the low-$z$ samples in the literature. 

\begin{figure}
    \centering
    \includegraphics[width=.48\textwidth]{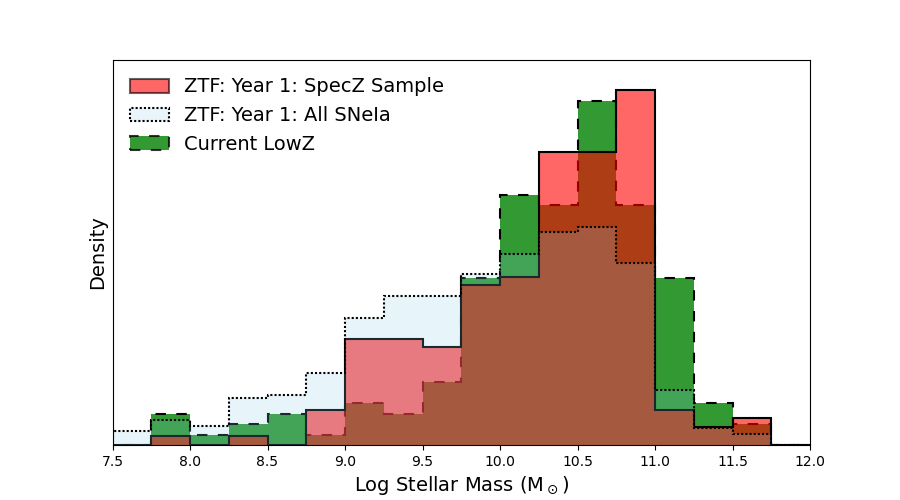}
    \caption{
   Log of the host stellar mass distribution for the DR1 host-$z$ sample (red),  DR1 all-$z$ sample (i.e. including SNe where the redshift is determined from the SN spectrum itself) (blue)  and the current low-$z$ SN~Ia sample  (green). The Y1 sample has $48.1 \%$ of all SNe~Ia in low mass host galaxies compared to 23.8$\%$ for the current low-$z$ distribution.}
    \label{fig:host_logmass}
\end{figure}

\subsubsection{Host galaxy mass}
Here we compute stellar masses ($M_{\rm stellar}$) for the host galaxies of SNe~Ia in our host-$z$ sample.
We take the host galaxy photometry from PanSTARRS legacy imaging \citep{2016arXiv161205560C} and identify the host galaxy using the \lq Directional Light Radius\rq\ \citep[DLR;][]{Sullivan2006,2012ApJ...755...61S} methodology, requiring DLR$<5$. SNe with no host satistfying this criteria (5; 2\%) are considered hostless. Stellar masses are computed by comparing measured \textit{griz}-band fluxes to those of spectral energy distribution (SED) model at the redshift of the SN. For this, we use the spectral synthesis code, P\'EGASE2 \citep{1997A&A...326..950F,2002A&A...386..446L} combined with a \citet{2001MNRAS.322..231K} initial mass function (IMF) to calculate synthetic SEDs. We consider 9 exponentially declining star formation histories (SFHs) evaluated at 102 time-steps with ${\rm SFR}(t) = {\rm exp} ^{-t/\tau} /\tau$, where $t$ is the age of the galaxy and $\tau$ is the e-folding time, each with 7 foreground dust screens   \citep[see][for more details]{2020MNRAS.494.4426S}. 
We present the $M_{\rm stellar}$ distribution for our sample in Figure~\ref{fig:host_logmass}. For comparison, we also analyse the host galaxies of the current low-$z$ sample with the same procedure. We find a slightly larger fraction of low (log $\frac{M_{\rm stellar}}{{\rm M}_{\odot}} < 10$ ) mass hosts compared to the literature sample. However, we note that requiring a spectroscopic redshift from the literature could lead to some preference against fainter and even lower mass hosts in the host-$z$ sample compared to the entire DR1 sample.  Therefore, we also compute the masses for the entire DR1 sample. We find that compared to the 23.8$\%$ of SNe~Ia in low mass galaxies in the current low-$z$, the ZTF Y1 spec-$z$ sample has 31.6$\%$ of the SNe~Ia in low mass galaxies and ZTF Y1 all-$z$ has 48.1$\%$ of the SNe~Ia in low mass galaxies.
A detailed analysis of the host galaxies of all SNe~Ia in the DR1 sample will be presented in a follow-up study (Smith et al. in preparation). This furthermore emphasizes the need for post-survey SN~Ia host spectroscopy to obtain redshifts.

\subsubsection{Hubble residuals}
For the sample after the selection cuts described above we derive the distribution of the Hubble residuals. From the fit to the magnitude-redshift relation, we derive the the rms ($\sigma_{\rm rms}$) and intrinsic scatter ($\sigma_{\rm int}$) values. The $\sigma_{\rm int}$ value is fitted simultaneously with the coefficient of the width-luminosity and colour luminosity relations, which allows us to robustly infer $\sigma_{\rm int}$, instead of fixing it  to give an adequate goodness of fit.

While for the lightcurve fitting we fix the SN model redshift to the heliocentric frame, for our inference of the Hubble residuals use CMB frame redshifts. We convert the heliocentric redshifts to CMB frame using the standard conversion formula and the dipole velocity and position of the CMB from \citet{1996ApJ...473..576F}. Since we cannot constrain $\Omega_{\rm M}$ from low-$z$ SNe~Ia themselves, we fix it to the value from \emph{Planck} of 0.307 \citep{2018arXiv180706209P}.

In our fit, the error propagated for each SN is the sum of the fit uncertainty from the SALT2 covariance matrix ($\sigma_{\rm fit}$), the peculiar velocity error ($\sigma_{\rm pec}$) and $\sigma_{\rm int}$. 
To get the error on the fit from the covariance matrix, we sample over the $\alpha$ and $\beta$ nuisance parameters along with the intercept of the Hubble diagram and the intrinsic scatter.
\begin{equation}
    \sigma^2_{\rm m} = \sigma^2_{\rm fit} + \sigma^2_{\rm pec} + \sigma^2_{\rm int}
    \label{eq:error_magnitude}
\end{equation}

The $\sigma_{\rm pec}$ magnitude error is derived assuming a stochastic peculiar velocity error value of 300 km\, s$^{-1}$ \citep{2015MNRAS.450..317C}, and $\sigma_{\rm int}$ to account for the intrinsic scatter of the SNe~Ia. Here, the $\sigma_{\rm fit}$ error term is derived from the output covariance matrix of the SALT2 model fit, for a given value of $\alpha$ and $\beta$. Hence, in our inference, we fit the magnitude redshift relation with four free parameters, namely, the intercept of the magnitude-redshift relation,  $\alpha$, $\beta$, $\sigma_{\rm int}$ with uninformative priors on each.  We use \texttt{PyMultiNest} \citep{2014A&A...564A.125B}, a python wrapper to \texttt{MultiNest} \citep{2009MNRAS.398.1601F} to derive the posterior distribution on the parameters.

For the best fit parameter estimates, we obtain a $\sigma_{\rm rms}$ of  0.17 mag. We note that this is comparable to the lightcurve scatter from other low-$z$ samples in the literature, when analysed with the SALT2 model \citep[e.g., see][]{2021arXiv210205678T} as well as low-$z$ sample of SNe~Ia used in cosmological studies \citep[e.g.][]{2018ApJ...859..101S}. 
Moreover, if we don't apply the final selection cut of the Chauvenet criterion and keep the 2 SNe~Ia that don't pass the cut in the sample, the $\sigma_{\rm rms} = 0.18$ mag, hence, the rms scatter is not impacted significantly by the removal of outliers. A comparison of the Hubble residuals with the current low-$z$ sample from the literature is shown in Figure~\ref{fig:salt_param_dist} (bottom panel). Our sample has a $\sigma_{\rm int} = 0.15$ mag.  We note that the rms dispersion and $\sigma_{\rm int}$ are  higher than report values for high-$z$ sample \citep[e.g.][]{2018ApJ...859..101S,2019ApJ...872L..30A}
We emphasise that the intrinsic scatter term is not derived for a complete treatment of the systematic uncertainties, but rather only includes a  statistical error term that is a combination of the fit error and the peculiar velocity error. We would expect the intrinsic scatter value to be lower when including a complete systematics uncertainty budget. 

\begin{figure}
    \centering
    \includegraphics[width=.5\textwidth, trim = 0 10 0 10]{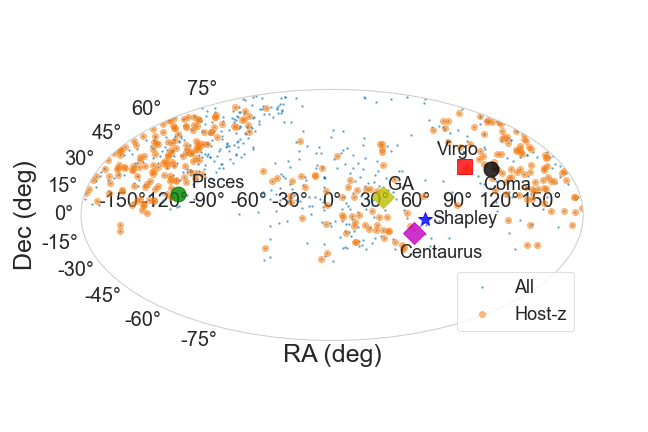}
    \caption{Sky distribution of the entire DR1 sample (DR1) with the subsample having host galaxy spec-z highlighted in orange. Known local overdensities, e.g. Coma (black circle), Virgo (red square), Great Attractor (yellow diamond), Shapley concentration (blue star), Pisces-Perseus cluster (green circle), Centaurus (magenta square) are highlighted for comparison.}
    \label{fig:skydist_clustercomp}
\end{figure}

We emphasize that since there are ongoing improvement to the data processing pipeline, the inferred root mean square (RMS) and intrinsic scatter values should be considered as preliminary upper limits, expected to improve in subsequent analyses. We caution against using the reported fit values for inferring cosmological parameters and conducting more detailed cosmological analyses, e.g. for $H_0$ or dark energy properties We have not yet produced a complete systematics error budget.
Therefore, at this stage in our analyses we are blind to the parameters governing the width-luminosity and colour luminosity relations as well as the intercept of the Hubble diagram till such a complete systematics analysis has been performed.

\subsection{Potential calibrator sample SNe~Ia}
\label{ssec:calib_samp}
In the sections above, we have focussed on the SNe~Ia at higher redshift in our sample. While the lowest redshift SNe in our dataset cannot be used for the Hubble flow rung of the distance ladder,  with future ground and spaced based observatories, we can expect measurements of independent, calibrated distances to their  host galaxies, using secondary distance indicators. This would be possible to distances of $D_{\rm L} \sim 80$ Mpc or approximately translating to $z \sim 0.02$ \citep[e.g., see][]{2019BAAS...51c.456B}. Such measurements in the literature use Cepheid variables \citep{riess2019}, TRGB stars \citep{Freedman2019}, JAGB stars \citep[for e.g.][]{2020ApJ...899...67F,2020arXiv201204536L} or Mira variables \citep{2020ApJ...889....5H}.

The large uncertainty from  heterogeneously calibrated instruments and photometric systems also impacts the local $H_0$ measurement via the calibrator sample of SNe~Ia. Hence, the ZTF sample offers a unique opportunity to have a homogeneously observed calibrator sample of SNe~Ia, on the same instrument as the Hubble flow SNe~Ia. 

In our sample, we have 14 SNe at $z < 0.02$. When applying the criteria for the final sample selection as in section~\ref{ssec:sample_dem}, we find that 6 SNe pass the cuts.  Of the SNe rejected by the cuts, there is one spectroscopically peculiar (91bg-like), three have high reddening ($c > 0.3$) and four do not have sufficient sampling near peak.  We emphasize that these six SNe are from the host-z sample. Since independent distances in the calibrator rung of the distance ladder do not require a precise redshift, for the complete DR1 sample we expect this nearby sample to $\sim$ twice as large, i.e. $\sim$ 10 SNe~Ia in the DR1 sample alone that we can get independent distances to the host galaxies in the near future.


ZTF-I has transitioned to  phase-II of operations (ZTF-II) from fall 2020. With ZTF-II ongoing, we expect to find a few tens of SNe~Ia in the redshift range $z < 0.02$ (ZTF-I + II combined) which would be a large sample, of order the current number of calibrator distances with all SNe on a single photometric system, which is the same as the system on which the Hubble flow SNe are observed.

\section{Discussion and Conclusions}
\label{sec:discussion}
We presented the first year dataset and results from the ZTF SN~Ia survey. The sample includes 761 spectroscopically confirmed SNe~Ia, 305 of which have spectroscopic redshifts from the host galaxy. The large sample statistics, early discovery and dense sampling make it a unique sample for cosmology with a uniformly measured low-$z$ SN~Ia sample. In the coming years, with the aid of multi-fibre spectroscopic surveys, we will complete the redshift measurements for the remaining sample of SN~Ia hosts. We note that the complete DR1 sample is $\sim$ factor of 3 larger than the current combined low-$z$ anchor and the host-$z$ subsample we present is by itself larger than the combined low-$z$ anchor as well.  ZTF has discovered $\sim$ 2.5 SNe~Ia per field, which, for an FoV of 47 deg$^{2}$, implies a discovery rate of $\sim$ 0.06 SNe deg$^{-2}$yr$^{-1}$, accounting for survey downtime in the first year of operations.

Our SN~Ia sample is created from an untargeted search and follow-up program. This specifically allows us to probe the underlying distribution of SN~Ia environment properties, which will be presented in companion studies. Hence, we can test the impact of SN~Ia luminosity - host galaxy correlations on the inferred cosmological parameters. This single system search and follow-up system matches very well with, and is complementary to, future SN cosmology programs e.g. Rubin Observatory \citep{2019ApJ...873..111I}, Roman Space Telescope \citep{2018ApJ...867...23H}, which are attuned to high-$z$ SNe~Ia. 

\begin{figure}
    \centering
    \includegraphics[width=.48\textwidth]{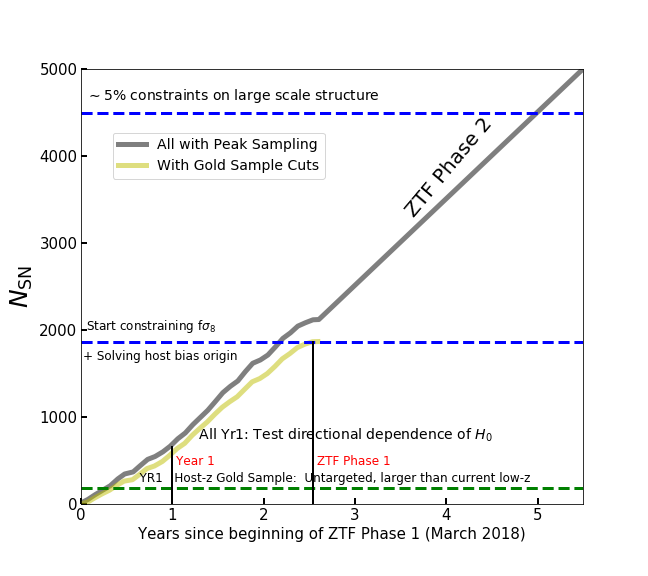}
    \caption{Number of SNe~Ia with sufficient data for measuring distances discovered by ZTF-I and expected to be discovered by the ongoing ZTF-II as a function of time since the beginning of ZTF-I. The vertical lines mark the DR1 sample (present here) and the total ZTF-I sample.}
    \label{fig:ztf1_2}
\end{figure}

We present well-sampled lightcurves for the 305 SNe~Ia in the host-$z$ sample.  From the multi-band data we derive lightcurve parameters and distance estimates. After applying selection cuts, we compute the Hubble residuals using 200 SNe that pass the cuts. This is already a competitive sample, discovered by an untargeted survey,  which is homogeneously observed on a single photometric system. The sample is also larger than the combined low-$z$ anchor sample used in the literature for cosmological constraints. Moreover, the median redshift of our sample is $\sim$ twice that of the current low-$z$ anchor sample. This is important to study local large scale structure are greater depths than with current samples.  Recent studies \citep[e.g.][]{feindt2013} look into constraining the origin of the local bulk flow with low-$z$ SNe~Ia. There are tentative suggestions that a combination of the Shapley concentration \citep{1930BHarO.874....9S,1991MNRAS.248..101R} and Sloan Great Wall could explain the size of dipole velocity (see Figure~\ref{fig:skydist_clustercomp} for positions of known local overdensities, overlaid on the SNe in our sample). With the current host-$z$ sample alone, in the highest redshift shell ($0.06 < z < 0.1$), which encompasses the Sloan Great Wall ($z \sim 0.07 - 0.08$) we have $\sim$ factor 2 improvement in sample statistics compared to the current SN~Ia compilations. With the entire DR1 sample, this can be an improvement of up to a factor of 4, important to test what is the origin of the local flow velocity.

 Our sample uniquely features very early discoveries, with a median first detection epoch of -13.5 days,  and densely sampled lightcurves. Recent studies in the literature \citep[e.g.][]{2019ApJ...871..219H} propose improved standardisation models using SN~Ia data on the rise. They split the canonical lightcurve shape parameter in SALT2 into $x_1^{\rm r}$ and  $x_1^{\rm f}$ for the rising and falling part of the lightcurve and find that the two lightcurve width model decreases the luminosity scatter. Our sample presented here is ideal for testing such novel standardisation procedures which will be explored in follow-up studies on improving lightcurve fitting algorithms.  


We presented the SALT2 $x_1$ and $c$ parameter distributions in Figure~\ref{fig:salt_param_dist}. We compared the distribution to the sample from the literature using a KS-test and found a low $p$-value for the $x_1$ distributions suggesting that they are not drawn from the same parent population. For the $c$-distribution we find that the $p$-value is high indicating that, for the $-0.3 < c < 0.3$ the distributions are drawn from the same parent population. However, we find a larger fraction of $c > 0.3$ SNe~Ia compared to the literature samples.  
We also present the stellar masses of the SN~Ia host galaxies. We find a slightly larger fraction of low $M_{\rm stellar}$ host galaxies compared to the literature sample. We note that our requirement of having a spectroscopic redshift of the host can introduce selection effects in our host-$z$ sample and will present a detail study of the entire DR1 host galaxy properties in a companion paper.
For our sample, we compute the Hubble residuals and find an rms dispersion of 0.17 mag, comparable to the current state-of-the-art in the literature. We emphasize that with ongoing improvements in the data processing this is expected to an upper limit on the scatter. We also simultaneously fit for $\sigma_{\rm int}$ using only the statistical errors from the lightcurve fit and peculiar velocity errors and find a value of 0.15 mag.


ZTF completed phase I of observations in October 2020, discovering and spectroscopically classifying $>$ 3000 SNe~Ia. 
In Figure~\ref{fig:ztf1_2} we present the number of SNe discovered as a function of the time at which they peak (relative to the beginning of ZTF-I). We present these values for the entire ZTF-I using the photometry from the alerts computed by IPAC \citep[see][for details]{2019PASP..131a8003M}. We restrict the sample with quality cuts to the lightcurve sampling as presented above and only show the SNe with an uncertainty on the inferred time of maximum of $<$ 1 day. The full DR1 sample of 761 SNe~Ia is already large enough we can start to test for anisotropies of the luminosity distance-redshift relation \citep[e.g.][]{2021arXiv210311918M}. 
The complete ZTF-I sample after the quality cuts on the number of lightcurve points and $\sigma(t_0)$ has $> 2400$ SNe~Ia. After applying the selection cuts for the gold sample based on SALT2 parameters, there are $> 1800$ SNe~Ia. We emphasise that these sample sizes are derived based on alert photometry which has very restrictive criteria and hence, with the photometry pipeline described in this work, we expect these numbers to improve further. With these large sample statistics, the already obtained ZTF-I data is competitve to measure local large scale structure properties, e.g. the growth rate, parametrised as $fD$ \citep[see][for more details]{graziani2020}. Analysing subsamples, based on host galaxy properties, from the complete ZTF-I dataset based on host galaxy properties will be critical to study the origin of environmental systematics in SN~Ia cosmology \citep{rigault2020,2018ApJ...867..108J}.

We have ongoing efforts to improve the data reduction pipeline, obtain host galaxy spectroscopic redshifts, quantify systematic uncertainties and finally constrain cosmological parameters. We expect the inferred intrinsic and rms dispersion values to decrease with improvements in the data processing. The size of the host-$z$ as well as the complete DR1 samples is already at a stage where, with the improvements mentioned above, we can improve constraints on dark energy, measure $H_0$ and the local bulk flow velocity. With the entire ZTF-I sample already acquired, we have sufficient number of SNe to obtain constraints on the growth of structure. This makes the ZTF dataset exciting for answering various questions in cosmology.

\label{sec:conclusions}
\section*{Acknowledgements}
S.D. is supported by the Isaac Newton Trust and the Kavli Foundation through the Newton-Kavli fellowship and acknowledges a research fellowship at Lucy Cavendish College. A.G acknowledges support from the Swedish Research Council under Dnr VR 2016-03274 and 2020-03444. M.S., M.R., and Y.-L.K. have received funding from the European Research Council (ERC) under the European Unions Horizon 2020 research and innovation program (grant agreement No. 759194 - USNAC). 

Based on observations obtained with the Samuel Oschin Telescope 48-inch and the 60-inch Telescope at the Palomar Observatory as part of the Zwicky Transient Facility project. ZTF is supported by the National Science Foundation under Grant No. AST-1440341 and a collaboration including Caltech, IPAC, the Weizmann Institute for Science, the Oskar Klein Center at Stockholm University, the University of Maryland, the University of Washington, Deutsches Elektronen-Synchrotron and Humboldt University, Los Alamos National Laboratories, the TANGO Consortium of Taiwan, the University of Wisconsin at Milwaukee, and Lawrence Berkeley National Laboratories. Operations are conducted by COO, IPAC, and UW.

SED Machine is based upon work supported by the National Science Foundation under Grant No. 1106171. The ZTF forced-photometry service was funded under the Heising-Simons Foundation grant. This work was supported by the
GROWTH project funded by the National Science Foundation under
Grant No 1545949. 

software: {Numpy~\citep{2011CSE....13b..22V}, Astropy~\citep{2013A&A...558A..33A,2018AJ....156..123A},~sncosmo~\citep{2016ascl.soft11017B}, PhotUtils~\citep{Bradley_2019_2533376}, ZTFQuery~\citep{2018zndo...1345222R}, SWARP~\citep{2010ascl.soft10068B}, HOTPANTS~\citep{2015ascl.soft04004B}, ZUDS, fringez~\citep{2021arXiv210210738M},  matplotlib~\citep{2007CSE.....9...90H}, IPAC forced Photometry Service, spextractor~\citep{Papadogiannakis1300923}.}
\section*{Data Availability}
The data used in this study are made available on github here: \url{https://github.com/ZwickyTransientFacility/ztfcosmodr}
 



\bibliographystyle{mnras}
\bibliography{ztfia} 

\begin{thebibliography}{}
\makeatletter
\relax
\def\mn@urlcharsother{\let\do\@makeother \do\$\do\&\do\#\do\^\do\_\do\%\do\~}
\def\mn@doi{\begingroup\mn@urlcharsother \@ifnextchar [ {\mn@doi@}
  {\mn@doi@[]}}
\def\mn@doi@[#1]#2{\def\@tempa{#1}\ifx\@tempa\@empty \href
  {http://dx.doi.org/#2} {doi:#2}\else \href {http://dx.doi.org/#2} {#1}\fi
  \endgroup}
\def\mn@eprint#1#2{\mn@eprint@#1:#2::\@nil}
\def\mn@eprint@arXiv#1{\href {http://arxiv.org/abs/#1} {{\tt arXiv:#1}}}
\def\mn@eprint@dblp#1{\href {http://dblp.uni-trier.de/rec/bibtex/#1.xml}
  {dblp:#1}}
\def\mn@eprint@#1:#2:#3:#4\@nil{\def\@tempa {#1}\def\@tempb {#2}\def\@tempc
  {#3}\ifx \@tempc \@empty \let \@tempc \@tempb \let \@tempb \@tempa \fi \ifx
  \@tempb \@empty \def\@tempb {arXiv}\fi \@ifundefined
  {mn@eprint@\@tempb}{\@tempb:\@tempc}{\expandafter \expandafter \csname
  mn@eprint@\@tempb\endcsname \expandafter{\@tempc}}}

\bibitem[\protect\citeauthoryear{{Abbott} et~al.,}{{Abbott}
  et~al.}{2019}]{2019ApJ...872L..30A}
{Abbott} T.~M.~C.,  et~al., 2019, \mn@doi [\apjl] {10.3847/2041-8213/ab04fa},
  \href {https://ui.adsabs.harvard.edu/abs/2019ApJ...872L..30A} {872, L30}

\bibitem[\protect\citeauthoryear{{Ahumada} et~al.,}{{Ahumada}
  et~al.}{2020}]{2020ApJS..249....3A}
{Ahumada} R.,  et~al., 2020, \mn@doi [\apjs] {10.3847/1538-4365/ab929e}, \href
  {https://ui.adsabs.harvard.edu/abs/2020ApJS..249....3A} {249, 3}

\bibitem[\protect\citeauthoryear{{Astropy Collaboration} et~al.,}{{Astropy
  Collaboration} et~al.}{2013}]{2013A&A...558A..33A}
{Astropy Collaboration} et~al., 2013, \mn@doi [\aap]
  {10.1051/0004-6361/201322068}, \href
  {https://ui.adsabs.harvard.edu/abs/2013A&A...558A..33A} {558, A33}

\bibitem[\protect\citeauthoryear{{Astropy Collaboration} et~al.,}{{Astropy
  Collaboration} et~al.}{2018}]{2018AJ....156..123A}
{Astropy Collaboration} et~al., 2018, \mn@doi [\aj] {10.3847/1538-3881/aabc4f},
  \href {https://ui.adsabs.harvard.edu/abs/2018AJ....156..123A} {156, 123}

\bibitem[\protect\citeauthoryear{{Barbary} et~al.,}{{Barbary}
  et~al.}{2016}]{2016ascl.soft11017B}
{Barbary} K.,  et~al., 2016, {SNCosmo: Python library for supernova cosmology}
  (\mn@eprint {ascl} {1611.017})

\bibitem[\protect\citeauthoryear{{Beaton} et~al.,}{{Beaton}
  et~al.}{2019}]{2019BAAS...51c.456B}
{Beaton} R.~L.,  et~al., 2019, \baas, \href
  {https://ui.adsabs.harvard.edu/abs/2019BAAS...51c.456B} {51, 456}

\bibitem[\protect\citeauthoryear{{Becker}}{{Becker}}{2015}]{2015ascl.soft04004B}
{Becker} A.,  2015, {HOTPANTS: High Order Transform of PSF ANd Template
  Subtraction} (\mn@eprint {ascl} {1504.004})

\bibitem[\protect\citeauthoryear{{Bellm} et~al.,}{{Bellm}
  et~al.}{2019a}]{bellm2019}
{Bellm} E.~C.,  et~al., 2019a, \mn@doi [\pasp] {10.1088/1538-3873/aaecbe},
  \href {https://ui.adsabs.harvard.edu/abs/2019PASP..131a8002B} {131, 018002}

\bibitem[\protect\citeauthoryear{{Bellm} et~al.,}{{Bellm}
  et~al.}{2019b}]{bellm2019scheduler}
{Bellm} E.~C.,  et~al., 2019b, \mn@doi [\pasp] {10.1088/1538-3873/ab0c2a},
  \href {https://ui.adsabs.harvard.edu/abs/2019PASP..131f8003B} {131, 068003}

\bibitem[\protect\citeauthoryear{{Ben-Ami}, {Konidaris}, {Quimby}, {Davis},
  {Ngeow}, {Ritter}  \& {Rudy}}{{Ben-Ami} et~al.}{2012}]{2012SPIE.8446E..86B}
{Ben-Ami} S.,  {Konidaris} N.,  {Quimby} R.,  {Davis} J.~T.,  {Ngeow} C.~C.,
  {Ritter} A.,   {Rudy} A.,  2012, in Ground-based and Airborne Instrumentation
  for Astronomy IV. p. 844686, \mn@doi{10.1117/12.926317}

\bibitem[\protect\citeauthoryear{{Bertin}}{{Bertin}}{2010}]{2010ascl.soft10068B}
{Bertin} E.,  2010, {SWarp: Resampling and Co-adding FITS Images Together}
  (\mn@eprint {ascl} {1010.068})

\bibitem[\protect\citeauthoryear{{Bertin}, {Mellier}, {Radovich}, {Missonnier},
  {Didelon}  \& {Morin}}{{Bertin} et~al.}{2002}]{2002ASPC..281..228B}
{Bertin} E.,  {Mellier} Y.,  {Radovich} M.,  {Missonnier} G.,  {Didelon} P.,
  {Morin} B.,  2002, in {Bohlender} D.~A.,  {Durand} D.,   {Handley} T.~H.,
  eds,  Astronomical Society of the Pacific Conference Series Vol. 281,
  Astronomical Data Analysis Software and Systems XI. p.~228

\bibitem[\protect\citeauthoryear{{Betoule} et~al.,}{{Betoule}
  et~al.}{2014}]{Betoule2014}
{Betoule} M.,  et~al., 2014, \mn@doi [\aap] {10.1051/0004-6361/201423413},
  \href {https://ui.adsabs.harvard.edu/abs/2014A&A...568A..22B} {568, A22}

\bibitem[\protect\citeauthoryear{{Blagorodnova} et~al.,}{{Blagorodnova}
  et~al.}{2018}]{2018PASP..130c5003B}
{Blagorodnova} N.,  et~al., 2018, \mn@doi [\pasp] {10.1088/1538-3873/aaa53f},
  \href {https://ui.adsabs.harvard.edu/abs/2018PASP..130c5003B} {130, 035003}

\bibitem[\protect\citeauthoryear{Bradley et~al.,}{Bradley
  et~al.}{2019}]{Bradley_2019_2533376}
Bradley L.,  et~al., 2019, astropy/photutils: v0.6,
  \mn@doi{10.5281/zenodo.2533376}, \url
  {https://doi.org/10.5281/zenodo.2533376}

\bibitem[\protect\citeauthoryear{Brout et~al.,}{Brout
  et~al.}{2019}]{Brout18-SMP}
Brout D.,  et~al., 2019, \mn@doi [The Astrophysical Journal]
  {10.3847/1538-4357/ab06c1}, 874, 106

\bibitem[\protect\citeauthoryear{{Buchner} et~al.,}{{Buchner}
  et~al.}{2014}]{2014A&A...564A.125B}
{Buchner} J.,  et~al., 2014, \mn@doi [\aap] {10.1051/0004-6361/201322971},
  \href {http://adsabs.harvard.edu/abs/2014A%26A...564A.125B} {564, A125}

\bibitem[\protect\citeauthoryear{{Bulla}, {Goobar}  \& {Dhawan}}{{Bulla}
  et~al.}{2018}]{2018MNRAS.479.3663B}
{Bulla} M.,  {Goobar} A.,   {Dhawan} S.,  2018, \mn@doi [\mnras]
  {10.1093/mnras/sty1619}, \href
  {https://ui.adsabs.harvard.edu/abs/2018MNRAS.479.3663B} {479, 3663}

\bibitem[\protect\citeauthoryear{{Bulla} et~al.,}{{Bulla}
  et~al.}{2020}]{2020ApJ...902...48B}
{Bulla} M.,  et~al., 2020, \mn@doi [\apj] {10.3847/1538-4357/abb13c}, \href
  {https://ui.adsabs.harvard.edu/abs/2020ApJ...902...48B} {902, 48}

\bibitem[\protect\citeauthoryear{{Burns} et~al.,}{{Burns}
  et~al.}{2011}]{burns2011}
{Burns} C.~R.,  et~al., 2011, \mn@doi [\aj] {10.1088/0004-6256/141/1/19}, \href
  {http://adsabs.harvard.edu/abs/2011AJ....141...19B} {141, 19}

\bibitem[\protect\citeauthoryear{{Burns} et~al.,}{{Burns}
  et~al.}{2018}]{2018ApJ...869...56B}
{Burns} C.~R.,  et~al., 2018, \mn@doi [\apj] {10.3847/1538-4357/aae51c}, \href
  {https://ui.adsabs.harvard.edu/abs/2018ApJ...869...56B} {869, 56}

\bibitem[\protect\citeauthoryear{{Cardelli}, {Clayton}  \& {Mathis}}{{Cardelli}
  et~al.}{1989}]{1989ApJ...345..245C}
{Cardelli} J.~A.,  {Clayton} G.~C.,   {Mathis} J.~S.,  1989, \mn@doi [\apj]
  {10.1086/167900}, \href
  {https://ui.adsabs.harvard.edu/abs/1989ApJ...345..245C} {345, 245}

\bibitem[\protect\citeauthoryear{{Carrick}, {Turnbull}, {Lavaux}  \&
  {Hudson}}{{Carrick} et~al.}{2015}]{2015MNRAS.450..317C}
{Carrick} J.,  {Turnbull} S.~J.,  {Lavaux} G.,   {Hudson} M.~J.,  2015, \mn@doi
  [\mnras] {10.1093/mnras/stv547}, \href
  {https://ui.adsabs.harvard.edu/abs/2015MNRAS.450..317C} {450, 317}

\bibitem[\protect\citeauthoryear{{Cenko} et~al.,}{{Cenko}
  et~al.}{2006}]{2006PASP..118.1396C}
{Cenko} S.~B.,  et~al., 2006, \mn@doi [\pasp] {10.1086/508366}, \href
  {https://ui.adsabs.harvard.edu/abs/2006PASP..118.1396C} {118, 1396}

\bibitem[\protect\citeauthoryear{{Chambers} et~al.,}{{Chambers}
  et~al.}{2016}]{2016arXiv161205560C}
{Chambers} K.~C.,  et~al., 2016, arXiv e-prints, \href
  {https://ui.adsabs.harvard.edu/abs/2016arXiv161205560C} {p. arXiv:1612.05560}

\bibitem[\protect\citeauthoryear{{Conley} et~al.,}{{Conley}
  et~al.}{2008}]{2008ApJ...681..482C}
{Conley} A.,  et~al., 2008, \mn@doi [\apj] {10.1086/588518}, \href
  {https://ui.adsabs.harvard.edu/abs/2008ApJ...681..482C} {681, 482}

\bibitem[\protect\citeauthoryear{{Conley} et~al.,}{{Conley}
  et~al.}{2011}]{2011ApJS..192....1C}
{Conley} A.,  et~al., 2011, \mn@doi [\apjs] {10.1088/0067-0049/192/1/1}, \href
  {https://ui.adsabs.harvard.edu/abs/2011ApJS..192....1C} {192, 1}

\bibitem[\protect\citeauthoryear{{DESI Collaboration} et~al.,}{{DESI
  Collaboration} et~al.}{2016}]{2016arXiv161100036D}
{DESI Collaboration} et~al., 2016, arXiv e-prints, \href
  {https://ui.adsabs.harvard.edu/abs/2016arXiv161100036D} {p. arXiv:1611.00036}

\bibitem[\protect\citeauthoryear{{Dekany} et~al.,}{{Dekany}
  et~al.}{2020}]{2020PASP..132c8001D}
{Dekany} R.,  et~al., 2020, \mn@doi [\pasp] {10.1088/1538-3873/ab4ca2}, \href
  {https://ui.adsabs.harvard.edu/abs/2020PASP..132c8001D} {132, 038001}

\bibitem[\protect\citeauthoryear{{Dettman} et~al.,}{{Dettman}
  et~al.}{2021}]{2021arXiv210206524D}
{Dettman} K.~G.,  et~al., 2021, arXiv e-prints, \href
  {https://ui.adsabs.harvard.edu/abs/2021arXiv210206524D} {p. arXiv:2102.06524}

\bibitem[\protect\citeauthoryear{{Dhawan}, {Goobar}, {M{\"o}rtsell},
  {Amanullah}  \& {Feindt}}{{Dhawan} et~al.}{2017}]{Dhawan2017b}
{Dhawan} S.,  {Goobar} A.,  {M{\"o}rtsell} E.,  {Amanullah} R.,   {Feindt} U.,
  2017, \mn@doi [\jcap] {10.1088/1475-7516/2017/07/040}, \href
  {https://ui.adsabs.harvard.edu/abs/2017JCAP...07..040D} {2017, 040}

\bibitem[\protect\citeauthoryear{{Feindt} et~al.,}{{Feindt}
  et~al.}{2013}]{feindt2013}
{Feindt} U.,  et~al., 2013, \mn@doi [\aap] {10.1051/0004-6361/201321880}, \href
  {https://ui.adsabs.harvard.edu/abs/2013A&A...560A..90F} {560, A90}

\bibitem[\protect\citeauthoryear{{Feindt}, {Nordin}, {Rigault}, {Brinnel},
  {Dhawan}, {Goobar}  \& {Kowalski}}{{Feindt}
  et~al.}{2019}]{2019JCAP...10..005F}
{Feindt} U.,  {Nordin} J.,  {Rigault} M.,  {Brinnel} V.,  {Dhawan} S.,
  {Goobar} A.,   {Kowalski} M.,  2019, \mn@doi [\jcap]
  {10.1088/1475-7516/2019/10/005}, \href
  {https://ui.adsabs.harvard.edu/abs/2019JCAP...10..005F} {2019, 005}

\bibitem[\protect\citeauthoryear{{Feroz}, {Hobson}  \& {Bridges}}{{Feroz}
  et~al.}{2009}]{2009MNRAS.398.1601F}
{Feroz} F.,  {Hobson} M.~P.,   {Bridges} M.,  2009, \mn@doi [\mnras]
  {10.1111/j.1365-2966.2009.14548.x}, \href
  {http://adsabs.harvard.edu/abs/2009MNRAS.398.1601F} {398, 1601}

\bibitem[\protect\citeauthoryear{{Filippenko} et~al.,}{{Filippenko}
  et~al.}{1992}]{1992AJ....104.1543F}
{Filippenko} A.~V.,  et~al., 1992, \mn@doi [\aj] {10.1086/116339}, \href
  {https://ui.adsabs.harvard.edu/abs/1992AJ....104.1543F} {104, 1543}

\bibitem[\protect\citeauthoryear{{Filippenko}, {Li}, {Treffers}  \&
  {Modjaz}}{{Filippenko} et~al.}{2001}]{2001ASPC..246..121F}
{Filippenko} A.~V.,  {Li} W.~D.,  {Treffers} R.~R.,   {Modjaz} M.,  2001, in
  {Paczynski} B.,  {Chen} W.-P.,   {Lemme} C.,  eds,  Astronomical Society of
  the Pacific Conference Series Vol. 246, IAU Colloq. 183: Small Telescope
  Astronomy on Global Scales. p.~121

\bibitem[\protect\citeauthoryear{{Fioc} \& {Rocca-Volmerange}}{{Fioc} \&
  {Rocca-Volmerange}}{1997}]{1997A&A...326..950F}
{Fioc} M.,  {Rocca-Volmerange} B.,  1997, \aap, \href
  {https://ui.adsabs.harvard.edu/abs/1997A&A...326..950F} {500, 507}

\bibitem[\protect\citeauthoryear{{Firth} et~al.,}{{Firth}
  et~al.}{2015}]{2015MNRAS.446.3895F}
{Firth} R.~E.,  et~al., 2015, \mn@doi [\mnras] {10.1093/mnras/stu2314}, \href
  {https://ui.adsabs.harvard.edu/abs/2015MNRAS.446.3895F} {446, 3895}

\bibitem[\protect\citeauthoryear{{Fixsen}, {Cheng}, {Gales}, {Mather}, {Shafer}
   \& {Wright}}{{Fixsen} et~al.}{1996}]{1996ApJ...473..576F}
{Fixsen} D.~J.,  {Cheng} E.~S.,  {Gales} J.~M.,  {Mather} J.~C.,  {Shafer}
  R.~A.,   {Wright} E.~L.,  1996, \mn@doi [\apj] {10.1086/178173}, \href
  {https://ui.adsabs.harvard.edu/abs/1996ApJ...473..576F} {473, 576}

\bibitem[\protect\citeauthoryear{{Folatelli} et~al.,}{{Folatelli}
  et~al.}{2010}]{2010AJ....139..120F}
{Folatelli} G.,  et~al., 2010, \mn@doi [\aj] {10.1088/0004-6256/139/1/120},
  \href {https://ui.adsabs.harvard.edu/abs/2010AJ....139..120F} {139, 120}

\bibitem[\protect\citeauthoryear{{Folatelli} et~al.,}{{Folatelli}
  et~al.}{2013}]{2013ApJ...773...53F}
{Folatelli} G.,  et~al., 2013, \mn@doi [\apj] {10.1088/0004-637X/773/1/53},
  \href {https://ui.adsabs.harvard.edu/abs/2013ApJ...773...53F} {773, 53}

\bibitem[\protect\citeauthoryear{{Foley} \& {Kasen}}{{Foley} \&
  {Kasen}}{2011}]{2011ApJ...729...55F}
{Foley} R.~J.,  {Kasen} D.,  2011, \mn@doi [\apj] {10.1088/0004-637X/729/1/55},
  \href {https://ui.adsabs.harvard.edu/abs/2011ApJ...729...55F} {729, 55}

\bibitem[\protect\citeauthoryear{{Foley} et~al.,}{{Foley}
  et~al.}{2013}]{2013ApJ...767...57F}
{Foley} R.~J.,  et~al., 2013, \mn@doi [\apj] {10.1088/0004-637X/767/1/57},
  \href {https://ui.adsabs.harvard.edu/abs/2013ApJ...767...57F} {767, 57}

\bibitem[\protect\citeauthoryear{{Foley} et~al.,}{{Foley}
  et~al.}{2018}]{foley2018}
{Foley} R.~J.,  et~al., 2018, \mn@doi [\mnras] {10.1093/mnras/stx3136}, \href
  {https://ui.adsabs.harvard.edu/abs/2018MNRAS.475..193F} {475, 193}

\bibitem[\protect\citeauthoryear{{Freedman}}{{Freedman}}{2021}]{2021arXiv210615656F}
{Freedman} W.~L.,  2021, arXiv e-prints, \href
  {https://ui.adsabs.harvard.edu/abs/2021arXiv210615656F} {p. arXiv:2106.15656}

\bibitem[\protect\citeauthoryear{{Freedman} \& {Madore}}{{Freedman} \&
  {Madore}}{2020}]{2020ApJ...899...67F}
{Freedman} W.~L.,  {Madore} B.~F.,  2020, \mn@doi [\apj]
  {10.3847/1538-4357/aba9d8}, \href
  {https://ui.adsabs.harvard.edu/abs/2020ApJ...899...67F} {899, 67}

\bibitem[\protect\citeauthoryear{{Freedman} et~al.,}{{Freedman}
  et~al.}{2009}]{2009ApJ...704.1036F}
{Freedman} W.~L.,  et~al., 2009, \mn@doi [\apj] {10.1088/0004-637X/704/2/1036},
  \href {https://ui.adsabs.harvard.edu/abs/2009ApJ...704.1036F} {704, 1036}

\bibitem[\protect\citeauthoryear{{Freedman} et~al.,}{{Freedman}
  et~al.}{2019}]{Freedman2019}
{Freedman} W.~L.,  et~al., 2019, \mn@doi [\apj] {10.3847/1538-4357/ab2f73},
  \href {https://ui.adsabs.harvard.edu/abs/2019ApJ...882...34F} {882, 34}

\bibitem[\protect\citeauthoryear{{Fremling} et~al.,}{{Fremling}
  et~al.}{2020}]{fremling2020}
{Fremling} C.,  et~al., 2020, \mn@doi [\apj] {10.3847/1538-4357/ab8943}, \href
  {https://ui.adsabs.harvard.edu/abs/2020ApJ...895...32F} {895, 32}

\bibitem[\protect\citeauthoryear{{GPy}}{{GPy}}{2012}]{gpy2014}
{GPy} since 2012, {GPy}: A Gaussian process framework in python,
  \url{http://github.com/SheffieldML/GPy}

\bibitem[\protect\citeauthoryear{{Ganeshalingam}, {Li}  \&
  {Filippenko}}{{Ganeshalingam} et~al.}{2011}]{2011MNRAS.416.2607G}
{Ganeshalingam} M.,  {Li} W.,   {Filippenko} A.~V.,  2011, \mn@doi [\mnras]
  {10.1111/j.1365-2966.2011.19213.x}, \href
  {https://ui.adsabs.harvard.edu/abs/2011MNRAS.416.2607G} {416, 2607}

\bibitem[\protect\citeauthoryear{{Goobar} \& {Leibundgut}}{{Goobar} \&
  {Leibundgut}}{2011}]{2011ARNPS..61..251G}
{Goobar} A.,  {Leibundgut} B.,  2011, \mn@doi [Annual Review of Nuclear and
  Particle Science] {10.1146/annurev-nucl-102010-130434}, \href
  {https://ui.adsabs.harvard.edu/abs/2011ARNPS..61..251G} {61, 251}

\bibitem[\protect\citeauthoryear{{Graham} et~al.,}{{Graham}
  et~al.}{2019}]{graham2019}
{Graham} M.~J.,  et~al., 2019, \mn@doi [\pasp] {10.1088/1538-3873/ab006c},
  \href {https://ui.adsabs.harvard.edu/abs/2019PASP..131g8001G} {131, 078001}

\bibitem[\protect\citeauthoryear{{Graziani} et~al.,}{{Graziani}
  et~al.}{2020}]{graziani2020}
{Graziani} R.,  et~al., 2020, arXiv e-prints, \href
  {https://ui.adsabs.harvard.edu/abs/2020arXiv200109095G} {p. arXiv:2001.09095}

\bibitem[\protect\citeauthoryear{{Guy}, {Astier}, {Nobili}, {Regnault}  \&
  {Pain}}{{Guy} et~al.}{2005}]{2005A&A...443..781G}
{Guy} J.,  {Astier} P.,  {Nobili} S.,  {Regnault} N.,   {Pain} R.,  2005,
  \mn@doi [\aap] {10.1051/0004-6361:20053025}, \href
  {https://ui.adsabs.harvard.edu/abs/2005A&A...443..781G} {443, 781}

\bibitem[\protect\citeauthoryear{{Guy} et~al.,}{{Guy} et~al.}{2007}]{guy2007}
{Guy} J.,  et~al., 2007, \mn@doi [\aap] {10.1051/0004-6361:20066930}, \href
  {http://adsabs.harvard.edu/abs/2007A%26A...466...11G} {466, 11}

\bibitem[\protect\citeauthoryear{{Guy} et~al.,}{{Guy} et~al.}{2010}]{guy2010}
{Guy} J.,  et~al., 2010, \mn@doi [\aap] {10.1051/0004-6361/201014468}, \href
  {https://ui.adsabs.harvard.edu/abs/2010A&A...523A...7G} {523, A7}

\bibitem[\protect\citeauthoryear{{Hamuy}, {Phillips}, {Suntzeff}, {Schommer},
  {Maza}, {Smith}, {Lira}  \& {Aviles}}{{Hamuy}
  et~al.}{1996}]{1996AJ....112.2438H}
{Hamuy} M.,  {Phillips} M.~M.,  {Suntzeff} N.~B.,  {Schommer} R.~A.,  {Maza}
  J.,  {Smith} R.~C.,  {Lira} P.,   {Aviles} R.,  1996, \mn@doi [\aj]
  {10.1086/118193}, \href
  {https://ui.adsabs.harvard.edu/abs/1996AJ....112.2438H} {112, 2438}

\bibitem[\protect\citeauthoryear{{Hamuy} et~al.,}{{Hamuy}
  et~al.}{2003}]{2003Natur.424..651H}
{Hamuy} M.,  et~al., 2003, \mn@doi [\nat] {10.1038/nature01854}, \href
  {https://ui.adsabs.harvard.edu/abs/2003Natur.424..651H} {424, 651}

\bibitem[\protect\citeauthoryear{{Hayden}, {Rubin}  \& {Strovink}}{{Hayden}
  et~al.}{2019}]{2019ApJ...871..219H}
{Hayden} B.,  {Rubin} D.,   {Strovink} M.,  2019, \mn@doi [\apj]
  {10.3847/1538-4357/aaf232}, \href
  {https://ui.adsabs.harvard.edu/abs/2019ApJ...871..219H} {871, 219}

\bibitem[\protect\citeauthoryear{{Hounsell} et~al.,}{{Hounsell}
  et~al.}{2018}]{2018ApJ...867...23H}
{Hounsell} R.,  et~al., 2018, \mn@doi [\apj] {10.3847/1538-4357/aac08b}, \href
  {https://ui.adsabs.harvard.edu/abs/2018ApJ...867...23H} {867, 23}

\bibitem[\protect\citeauthoryear{{Hsiao} et~al.,}{{Hsiao}
  et~al.}{2015}]{2015A&A...578A...9H}
{Hsiao} E.~Y.,  et~al., 2015, \mn@doi [\aap] {10.1051/0004-6361/201425297},
  \href {https://ui.adsabs.harvard.edu/abs/2015A&A...578A...9H} {578, A9}

\bibitem[\protect\citeauthoryear{{Huang} et~al.,}{{Huang}
  et~al.}{2020}]{2020ApJ...889....5H}
{Huang} C.~D.,  et~al., 2020, \mn@doi [\apj] {10.3847/1538-4357/ab5dbd}, \href
  {https://ui.adsabs.harvard.edu/abs/2020ApJ...889....5H} {889, 5}

\bibitem[\protect\citeauthoryear{{Hunter}}{{Hunter}}{2007}]{2007CSE.....9...90H}
{Hunter} J.~D.,  2007, \mn@doi [Computing in Science and Engineering]
  {10.1109/MCSE.2007.55}, \href
  {https://ui.adsabs.harvard.edu/abs/2007CSE.....9...90H} {9, 90}

\bibitem[\protect\citeauthoryear{{Huterer}}{{Huterer}}{2020}]{2020arXiv201005765H}
{Huterer} D.,  2020, arXiv e-prints, \href
  {https://ui.adsabs.harvard.edu/abs/2020arXiv201005765H} {p. arXiv:2010.05765}

\bibitem[\protect\citeauthoryear{{Huterer}, {Shafer}, {Scolnic}  \&
  {Schmidt}}{{Huterer} et~al.}{2017}]{2017JCAP...05..015H}
{Huterer} D.,  {Shafer} D.~L.,  {Scolnic} D.~M.,   {Schmidt} F.,  2017, \mn@doi
  [\jcap] {10.1088/1475-7516/2017/05/015}, \href
  {https://ui.adsabs.harvard.edu/abs/2017JCAP...05..015H} {2017, 015}

\bibitem[\protect\citeauthoryear{{Ivezi{\'c}} et~al.,}{{Ivezi{\'c}}
  et~al.}{2019}]{2019ApJ...873..111I}
{Ivezi{\'c}} {\v{Z}}.,  et~al., 2019, \mn@doi [\apj]
  {10.3847/1538-4357/ab042c}, \href
  {https://ui.adsabs.harvard.edu/abs/2019ApJ...873..111I} {873, 111}

\bibitem[\protect\citeauthoryear{{Jha}, {Branch}, {Chornock}, {Foley}, {Li},
  {Swift}, {Casebeer}  \& {Filippenko}}{{Jha}
  et~al.}{2006}]{2006AJ....132..189J}
{Jha} S.,  {Branch} D.,  {Chornock} R.,  {Foley} R.~J.,  {Li} W.,  {Swift}
  B.~J.,  {Casebeer} D.,   {Filippenko} A.~V.,  2006, \mn@doi [\aj]
  {10.1086/504599}, \href
  {https://ui.adsabs.harvard.edu/abs/2006AJ....132..189J} {132, 189}

\bibitem[\protect\citeauthoryear{{Jha}, {Riess}  \& {Kirshner}}{{Jha}
  et~al.}{2007}]{2007ApJ...659..122J}
{Jha} S.,  {Riess} A.~G.,   {Kirshner} R.~P.,  2007, \mn@doi [\apj]
  {10.1086/512054}, \href
  {https://ui.adsabs.harvard.edu/abs/2007ApJ...659..122J} {659, 122}

\bibitem[\protect\citeauthoryear{Jones et~al.,}{Jones
  et~al.}{2018a}]{Jones_2018}
Jones D.~O.,  et~al., 2018a, \mn@doi [The Astrophysical Journal]
  {10.3847/1538-4357/aae2b9}, 867, 108

\bibitem[\protect\citeauthoryear{{Jones} et~al.,}{{Jones}
  et~al.}{2018b}]{2018ApJ...867..108J}
{Jones} D.~O.,  et~al., 2018b, \mn@doi [\apj] {10.3847/1538-4357/aae2b9}, \href
  {https://ui.adsabs.harvard.edu/abs/2018ApJ...867..108J} {867, 108}

\bibitem[\protect\citeauthoryear{{Jones} et~al.,}{{Jones}
  et~al.}{2019}]{2019ApJ...881...19J}
{Jones} D.~O.,  et~al., 2019, \mn@doi [\apj] {10.3847/1538-4357/ab2bec}, \href
  {https://ui.adsabs.harvard.edu/abs/2019ApJ...881...19J} {881, 19}

\bibitem[\protect\citeauthoryear{{Jones} et~al.,}{{Jones}
  et~al.}{2021}]{2021ApJ...908..143J}
{Jones} D.~O.,  et~al., 2021, \mn@doi [\apj] {10.3847/1538-4357/abd7f5}, \href
  {https://ui.adsabs.harvard.edu/abs/2021ApJ...908..143J} {908, 143}

\bibitem[\protect\citeauthoryear{{Kasliwal} et~al.,}{{Kasliwal}
  et~al.}{2019}]{2019PASP..131c8003K}
{Kasliwal} M.~M.,  et~al., 2019, \mn@doi [\pasp] {10.1088/1538-3873/aafbc2},
  \href {http://adsabs.harvard.edu/abs/2019PASP..131c8003K} {131, 038003}

\bibitem[\protect\citeauthoryear{{Kessler} et~al.,}{{Kessler}
  et~al.}{2009}]{kessler09a}
{Kessler} R.,  et~al., 2009, \mn@doi [\apjs] {10.1088/0067-0049/185/1/32},
  \href {https://ui.adsabs.harvard.edu/abs/2009ApJS..185...32K} {185, 32}

\bibitem[\protect\citeauthoryear{{Knox} \& {Millea}}{{Knox} \&
  {Millea}}{2020}]{2020PhRvD.101d3533K}
{Knox} L.,  {Millea} M.,  2020, \mn@doi [\prd] {10.1103/PhysRevD.101.043533},
  \href {https://ui.adsabs.harvard.edu/abs/2020PhRvD.101d3533K} {101, 043533}

\bibitem[\protect\citeauthoryear{{Kourkchi} et~al.,}{{Kourkchi}
  et~al.}{2020}]{kourkchi2020}
{Kourkchi} E.,  et~al., 2020, \mn@doi [\apj] {10.3847/1538-4357/abb66b}, \href
  {https://ui.adsabs.harvard.edu/abs/2020ApJ...902..145K} {902, 145}

\bibitem[\protect\citeauthoryear{{Kroupa}}{{Kroupa}}{2001}]{2001MNRAS.322..231K}
{Kroupa} P.,  2001, \mn@doi [\mnras] {10.1046/j.1365-8711.2001.04022.x}, \href
  {https://ui.adsabs.harvard.edu/abs/2001MNRAS.322..231K} {322, 231}

\bibitem[\protect\citeauthoryear{{Le Borgne} \& {Rocca-Volmerange}}{{Le Borgne}
  \& {Rocca-Volmerange}}{2002}]{2002A&A...386..446L}
{Le Borgne} D.,  {Rocca-Volmerange} B.,  2002, \mn@doi [\aap]
  {10.1051/0004-6361:20020259}, \href
  {https://ui.adsabs.harvard.edu/abs/2002A&A...386..446L} {386, 446}

\bibitem[\protect\citeauthoryear{{Lee}, {Freedman}, {Madore}, {Owens}, {Monson}
   \& {Hoyt}}{{Lee} et~al.}{2020}]{2020arXiv201204536L}
{Lee} A.~J.,  {Freedman} W.~L.,  {Madore} B.~F.,  {Owens} K.~A.,  {Monson}
  A.~J.,   {Hoyt} T.~J.,  2020, arXiv e-prints, \href
  {https://ui.adsabs.harvard.edu/abs/2020arXiv201204536L} {p. arXiv:2012.04536}

\bibitem[\protect\citeauthoryear{{Leibundgut} \& {Sullivan}}{{Leibundgut} \&
  {Sullivan}}{2018}]{2018SSRv..214...57L}
{Leibundgut} B.,  {Sullivan} M.,  2018, \mn@doi [\ssr]
  {10.1007/s11214-018-0491-8}, \href
  {https://ui.adsabs.harvard.edu/abs/2018SSRv..214...57L} {214, 57}

\bibitem[\protect\citeauthoryear{{Leibundgut} et~al.,}{{Leibundgut}
  et~al.}{1993}]{leibundgut1993}
{Leibundgut} B.,  et~al., 1993, \mn@doi [\aj] {10.1086/116427}, \href
  {http://adsabs.harvard.edu/abs/1993AJ....105..301L} {105, 301}

\bibitem[\protect\citeauthoryear{{Lonappan}, {Kumar}, {Ruchika}, {Dinda}  \&
  {Sen}}{{Lonappan} et~al.}{2018}]{2018PhRvD..97d3524L}
{Lonappan} A.~I.,  {Kumar} S.,  {Ruchika} {Dinda} B.~R.,   {Sen} A.~A.,  2018,
  \mn@doi [\prd] {10.1103/PhysRevD.97.043524}, \href
  {https://ui.adsabs.harvard.edu/abs/2018PhRvD..97d3524L} {97, 043524}

\bibitem[\protect\citeauthoryear{{Macpherson} \& {Heinesen}}{{Macpherson} \&
  {Heinesen}}{2021}]{2021arXiv210311918M}
{Macpherson} H.~J.,  {Heinesen} A.,  2021, arXiv e-prints, \href
  {https://ui.adsabs.harvard.edu/abs/2021arXiv210311918M} {p. arXiv:2103.11918}

\bibitem[\protect\citeauthoryear{{Mandel}, {Wood-Vasey}, {Friedman}  \&
  {Kirshner}}{{Mandel} et~al.}{2009}]{2009ApJ...704..629M}
{Mandel} K.~S.,  {Wood-Vasey} W.~M.,  {Friedman} A.~S.,   {Kirshner} R.~P.,
  2009, \mn@doi [\apj] {10.1088/0004-637X/704/1/629}, \href
  {https://ui.adsabs.harvard.edu/abs/2009ApJ...704..629M} {704, 629}

\bibitem[\protect\citeauthoryear{{Mandel}, {Narayan}  \& {Kirshner}}{{Mandel}
  et~al.}{2011}]{mandel2011}
{Mandel} K.~S.,  {Narayan} G.,   {Kirshner} R.~P.,  2011, \mn@doi [\apj]
  {10.1088/0004-637X/731/2/120}, \href
  {http://adsabs.harvard.edu/abs/2011ApJ...731..120M} {731, 120}

\bibitem[\protect\citeauthoryear{{Mandel}, {Thorp}, {Narayan}, {Friedman}  \&
  {Avelino}}{{Mandel} et~al.}{2020}]{2020arXiv200807538M}
{Mandel} K.~S.,  {Thorp} S.,  {Narayan} G.,  {Friedman} A.~S.,   {Avelino} A.,
  2020, arXiv e-prints, \href
  {https://ui.adsabs.harvard.edu/abs/2020arXiv200807538M} {p. arXiv:2008.07538}

\bibitem[\protect\citeauthoryear{{Masci} et~al.,}{{Masci}
  et~al.}{2019}]{2019PASP..131a8003M}
{Masci} F.~J.,  et~al., 2019, \mn@doi [\pasp] {10.1088/1538-3873/aae8ac}, \href
  {https://ui.adsabs.harvard.edu/abs/2019PASP..131a8003M} {131, 018003}

\bibitem[\protect\citeauthoryear{{Mathews}, {Rose}, {Garnavich}, {Yamazaki}  \&
  {Kajino}}{{Mathews} et~al.}{2016}]{2016ApJ...827...60M}
{Mathews} G.~J.,  {Rose} B.~M.,  {Garnavich} P.~M.,  {Yamazaki} D.~G.,
  {Kajino} T.,  2016, \mn@doi [\apj] {10.3847/0004-637X/827/1/60}, \href
  {https://ui.adsabs.harvard.edu/abs/2016ApJ...827...60M} {827, 60}

\bibitem[\protect\citeauthoryear{{Medford} et~al.,}{{Medford}
  et~al.}{2021}]{2021arXiv210210738M}
{Medford} M.~S.,  et~al., 2021, arXiv e-prints, \href
  {https://ui.adsabs.harvard.edu/abs/2021arXiv210210738M} {p. arXiv:2102.10738}

\bibitem[\protect\citeauthoryear{{Miller} et~al.,}{{Miller}
  et~al.}{2020}]{2020ApJ...902...47M}
{Miller} A.~A.,  et~al., 2020, \mn@doi [\apj] {10.3847/1538-4357/abb13b}, \href
  {https://ui.adsabs.harvard.edu/abs/2020ApJ...902...47M} {902, 47}

\bibitem[\protect\citeauthoryear{{Mortsell}, {Goobar}, {Johansson}  \&
  {Dhawan}}{{Mortsell} et~al.}{2021a}]{2021arXiv210511461M}
{Mortsell} E.,  {Goobar} A.,  {Johansson} J.,   {Dhawan} S.,  2021a, arXiv
  e-prints, \href {https://ui.adsabs.harvard.edu/abs/2021arXiv210511461M} {p.
  arXiv:2105.11461}

\bibitem[\protect\citeauthoryear{{Mortsell}, {Goobar}, {Johansson}  \&
  {Dhawan}}{{Mortsell} et~al.}{2021b}]{2021arXiv210609400M}
{Mortsell} E.,  {Goobar} A.,  {Johansson} J.,   {Dhawan} S.,  2021b, arXiv
  e-prints, \href {https://ui.adsabs.harvard.edu/abs/2021arXiv210609400M} {p.
  arXiv:2106.09400}

\bibitem[\protect\citeauthoryear{{Nobili} et~al.,}{{Nobili}
  et~al.}{2005}]{2005A&A...437..789N}
{Nobili} S.,  et~al., 2005, \mn@doi [\aap] {10.1051/0004-6361:20042463}, \href
  {https://ui.adsabs.harvard.edu/abs/2005A&A...437..789N} {437, 789}

\bibitem[\protect\citeauthoryear{Papadogiannakis}{Papadogiannakis}{2019}]{Papadogiannakis1300923}
Papadogiannakis S.,  2019, PhD thesis, Stockholm University, Department of
  Physics

\bibitem[\protect\citeauthoryear{{Perley} et~al.,}{{Perley}
  et~al.}{2020}]{2020ApJ...904...35P}
{Perley} D.~A.,  et~al., 2020, \mn@doi [\apj] {10.3847/1538-4357/abbd98}, \href
  {https://ui.adsabs.harvard.edu/abs/2020ApJ...904...35P} {904, 35}

\bibitem[\protect\citeauthoryear{{Perlmutter} et~al.,}{{Perlmutter}
  et~al.}{1999}]{Perlmutter:1998np}
{Perlmutter} S.,  et~al., 1999, \mn@doi [\apj] {10.1086/307221}, \href
  {https://ui.adsabs.harvard.edu/abs/1999ApJ...517..565P} {517, 565}

\bibitem[\protect\citeauthoryear{{Phillips}}{{Phillips}}{1993}]{phillips1993}
{Phillips} M.~M.,  1993, \mn@doi [\apjl] {10.1086/186970}, \href
  {http://adsabs.harvard.edu/abs/1993ApJ...413L.105P} {413, L105}

\bibitem[\protect\citeauthoryear{{Planck Collaboration} et~al.,}{{Planck
  Collaboration} et~al.}{2018}]{2018arXiv180706209P}
{Planck Collaboration} et~al., 2018, arXiv e-prints, \href
  {https://ui.adsabs.harvard.edu/abs/2018arXiv180706209P} {p. arXiv:1807.06209}

\bibitem[\protect\citeauthoryear{Rasmussen \& Williams}{Rasmussen \&
  Williams}{2005}]{Rasmussen:2005:GPM:1162254}
Rasmussen C.~E.,  Williams C. K.~I.,  2005, Gaussian Processes for Machine
  Learning (Adaptive Computation and Machine Learning).
The MIT Press

\bibitem[\protect\citeauthoryear{{Rau} et~al.,}{{Rau}
  et~al.}{2009}]{2009PASP..121.1334R}
{Rau} A.,  et~al., 2009, \mn@doi [\pasp] {10.1086/605911}, \href
  {https://ui.adsabs.harvard.edu/abs/2009PASP..121.1334R} {121, 1334}

\bibitem[\protect\citeauthoryear{{Raychaudhury}, {Fabian}, {Edge}, {Jones}  \&
  {Forman}}{{Raychaudhury} et~al.}{1991}]{1991MNRAS.248..101R}
{Raychaudhury} S.,  {Fabian} A.~C.,  {Edge} A.~C.,  {Jones} C.,   {Forman} W.,
  1991, \mn@doi [\mnras] {10.1093/mnras/248.1.101}, \href
  {https://ui.adsabs.harvard.edu/abs/1991MNRAS.248..101R} {248, 101}

\bibitem[\protect\citeauthoryear{Rest et~al.,}{Rest et~al.}{2014}]{rest14}
Rest A.,  et~al., 2014, \mn@doi [The Astrophysical Journal]
  {10.1088/0004-637x/795/1/44}, 795, 44

\bibitem[\protect\citeauthoryear{{Riess} et~al.,}{{Riess}
  et~al.}{1998}]{Riess:1998cb}
{Riess} A.~G.,  et~al., 1998, \mn@doi [\aj] {10.1086/300499}, \href
  {https://ui.adsabs.harvard.edu/abs/1998AJ....116.1009R} {116, 1009}

\bibitem[\protect\citeauthoryear{{Riess}, {Casertano}, {Yuan}, {Macri}  \&
  {Scolnic}}{{Riess} et~al.}{2019a}]{2019ApJ...876...85R}
{Riess} A.~G.,  {Casertano} S.,  {Yuan} W.,  {Macri} L.~M.,   {Scolnic} D.,
  2019a, \mn@doi [\apj] {10.3847/1538-4357/ab1422}, \href
  {https://ui.adsabs.harvard.edu/abs/2019ApJ...876...85R} {876, 85}

\bibitem[\protect\citeauthoryear{{Riess}, {Casertano}, {Yuan}, {Macri}  \&
  {Scolnic}}{{Riess} et~al.}{2019b}]{riess2019}
{Riess} A.~G.,  {Casertano} S.,  {Yuan} W.,  {Macri} L.~M.,   {Scolnic} D.,
  2019b, \mn@doi [\apj] {10.3847/1538-4357/ab1422}, \href
  {https://ui.adsabs.harvard.edu/abs/2019ApJ...876...85R} {876, 85}

\bibitem[\protect\citeauthoryear{{Rigault}}{{Rigault}}{2018}]{2018zndo...1345222R}
{Rigault} M.,  2018, {ztfquery, a python tool to access ZTF data},
  \mn@doi{10.5281/zenodo.1345222}

\bibitem[\protect\citeauthoryear{{Rigault} et~al.,}{{Rigault}
  et~al.}{2015}]{rigault2015}
{Rigault} M.,  et~al., 2015, \mn@doi [\apj] {10.1088/0004-637X/802/1/20}, \href
  {https://ui.adsabs.harvard.edu/abs/2015ApJ...802...20R} {802, 20}

\bibitem[\protect\citeauthoryear{{Rigault} et~al.,}{{Rigault}
  et~al.}{2019}]{2019A&A...627A.115R}
{Rigault} M.,  et~al., 2019, \mn@doi [\aap] {10.1051/0004-6361/201935344},
  \href {https://ui.adsabs.harvard.edu/abs/2019A&A...627A.115R} {627, A115}

\bibitem[\protect\citeauthoryear{{Rigault} et~al.,}{{Rigault}
  et~al.}{2020}]{rigault2020}
{Rigault} M.,  et~al., 2020, \mn@doi [\aap] {10.1051/0004-6361/201730404},
  \href {https://ui.adsabs.harvard.edu/abs/2020A&A...644A.176R} {644, A176}

\bibitem[\protect\citeauthoryear{{Schlafly} \& {Finkbeiner}}{{Schlafly} \&
  {Finkbeiner}}{2011}]{2011ApJ...737..103S}
{Schlafly} E.~F.,  {Finkbeiner} D.~P.,  2011, \mn@doi [\apj]
  {10.1088/0004-637X/737/2/103}, \href
  {https://ui.adsabs.harvard.edu/abs/2011ApJ...737..103S} {737, 103}

\bibitem[\protect\citeauthoryear{{Schlegel}, {Finkbeiner}  \&
  {Davis}}{{Schlegel} et~al.}{1998}]{1998ApJ...500..525S}
{Schlegel} D.~J.,  {Finkbeiner} D.~P.,   {Davis} M.,  1998, \mn@doi [\apj]
  {10.1086/305772}, \href
  {https://ui.adsabs.harvard.edu/abs/1998ApJ...500..525S} {500, 525}

\bibitem[\protect\citeauthoryear{{Scolnic}, {Riess}, {Foley}, {Rest}, {Rodney},
  {Brout}  \& {Jones}}{{Scolnic} et~al.}{2014}]{2014ApJ...780...37S}
{Scolnic} D.~M.,  {Riess} A.~G.,  {Foley} R.~J.,  {Rest} A.,  {Rodney} S.~A.,
  {Brout} D.~J.,   {Jones} D.~O.,  2014, \mn@doi [\apj]
  {10.1088/0004-637X/780/1/37}, \href
  {https://ui.adsabs.harvard.edu/abs/2014ApJ...780...37S} {780, 37}

\bibitem[\protect\citeauthoryear{{Scolnic} et~al.,}{{Scolnic}
  et~al.}{2018}]{2018ApJ...859..101S}
{Scolnic} D.~M.,  et~al., 2018, \mn@doi [\apj] {10.3847/1538-4357/aab9bb},
  \href {https://ui.adsabs.harvard.edu/abs/2018ApJ...859..101S} {859, 101}

\bibitem[\protect\citeauthoryear{{Scolnic} et~al.,}{{Scolnic}
  et~al.}{2019}]{scolnic2019}
{Scolnic} D.,  et~al., 2019, Astro2020: Decadal Survey on Astronomy and
  Astrophysics, \href {https://ui.adsabs.harvard.edu/abs/2019astro2020T.270S}
  {2020, 270}

\bibitem[\protect\citeauthoryear{{Shapley}}{{Shapley}}{1930}]{1930BHarO.874....9S}
{Shapley} H.,  1930, Harvard College Observatory Bulletin, \href
  {https://ui.adsabs.harvard.edu/abs/1930BHarO.874....9S} {874, 9}

\bibitem[\protect\citeauthoryear{{Shappee} et~al.,}{{Shappee}
  et~al.}{2014}]{2014ApJ...788...48S}
{Shappee} B.~J.,  et~al., 2014, \mn@doi [\apj] {10.1088/0004-637X/788/1/48},
  \href {https://ui.adsabs.harvard.edu/abs/2014ApJ...788...48S} {788, 48}

\bibitem[\protect\citeauthoryear{{Shariff}, {Dhawan}, {Jiao}, {Leibundgut},
  {Trotta}  \& {van Dyk}}{{Shariff} et~al.}{2016}]{2016MNRAS.463.4311S}
{Shariff} H.,  {Dhawan} S.,  {Jiao} X.,  {Leibundgut} B.,  {Trotta} R.,   {van
  Dyk} D.~A.,  2016, \mn@doi [\mnras] {10.1093/mnras/stw2278}, \href
  {https://ui.adsabs.harvard.edu/abs/2016MNRAS.463.4311S} {463, 4311}

\bibitem[\protect\citeauthoryear{{Siebert}, {Foley}, {Jones}  \&
  {Davis}}{{Siebert} et~al.}{2020}]{2020MNRAS.493.5713S}
{Siebert} M.~R.,  {Foley} R.~J.,  {Jones} D.~O.,   {Davis} K.~W.,  2020,
  \mn@doi [\mnras] {10.1093/mnras/staa577}, \href
  {https://ui.adsabs.harvard.edu/abs/2020MNRAS.493.5713S} {493, 5713}

\bibitem[\protect\citeauthoryear{{Smith} et~al.,}{{Smith}
  et~al.}{2012}]{2012ApJ...755...61S}
{Smith} M.,  et~al., 2012, \mn@doi [\apj] {10.1088/0004-637X/755/1/61}, \href
  {https://ui.adsabs.harvard.edu/abs/2012ApJ...755...61S} {755, 61}

\bibitem[\protect\citeauthoryear{{Smith} et~al.,}{{Smith}
  et~al.}{2020}]{2020MNRAS.494.4426S}
{Smith} M.,  et~al., 2020, \mn@doi [\mnras] {10.1093/mnras/staa946}, \href
  {https://ui.adsabs.harvard.edu/abs/2020MNRAS.494.4426S} {494, 4426}

\bibitem[\protect\citeauthoryear{{Sullivan} et~al.,}{{Sullivan}
  et~al.}{2006}]{Sullivan2006}
{Sullivan} M.,  et~al., 2006, \mn@doi [\apj] {10.1086/506137}, \href
  {https://ui.adsabs.harvard.edu/abs/2006ApJ...648..868S} {648, 868}

\bibitem[\protect\citeauthoryear{Sullivan et~al.,}{Sullivan
  et~al.}{2011}]{Sullivan_2011}
Sullivan M.,  et~al., 2011, \mn@doi [The Astrophysical Journal]
  {10.1088/0004-637x/737/2/102}, 737, 102

\bibitem[\protect\citeauthoryear{{Swann} et~al.,}{{Swann}
  et~al.}{2019}]{2019Msngr.175...58S}
{Swann} E.,  et~al., 2019, \mn@doi [The Messenger] {10.18727/0722-6691/5129},
  \href {https://ui.adsabs.harvard.edu/abs/2019Msngr.175...58S} {175, 58}

\bibitem[\protect\citeauthoryear{{Taubenberger}}{{Taubenberger}}{2017}]{2017hsn..book..317T}
{Taubenberger} S.,  2017, {The Extremes of Thermonuclear Supernovae}.
p.~317, \mn@doi{10.1007/978-3-319-21846-5_37}

\bibitem[\protect\citeauthoryear{{The LSST Dark Energy Science Collaboration}
  et~al.,}{{The LSST Dark Energy Science Collaboration}
  et~al.}{2018}]{2018arXiv180901669T}
{The LSST Dark Energy Science Collaboration} et~al., 2018, arXiv e-prints,
  \href {https://ui.adsabs.harvard.edu/abs/2018arXiv180901669T} {p.
  arXiv:1809.01669}

\bibitem[\protect\citeauthoryear{{Thorp}, {Mandel}, {Jones}, {Ward}  \&
  {Narayan}}{{Thorp} et~al.}{2021}]{2021arXiv210205678T}
{Thorp} S.,  {Mandel} K.~S.,  {Jones} D.~O.,  {Ward} S.~M.,   {Narayan} G.,
  2021, arXiv e-prints, \href
  {https://ui.adsabs.harvard.edu/abs/2021arXiv210205678T} {p. arXiv:2102.05678}

\bibitem[\protect\citeauthoryear{{Tonry} et~al.,}{{Tonry}
  et~al.}{2018}]{2018PASP..130f4505T}
{Tonry} J.~L.,  et~al., 2018, \mn@doi [\pasp] {10.1088/1538-3873/aabadf}, \href
  {https://ui.adsabs.harvard.edu/abs/2018PASP..130f4505T} {130, 064505}

\bibitem[\protect\citeauthoryear{{Tripp}}{{Tripp}}{1998}]{tripp1998}
{Tripp} R.,  1998, \aap, \href
  {http://adsabs.harvard.edu/abs/1998A%26A...331..815T} {331, 815}

\bibitem[\protect\citeauthoryear{{Wang} et~al.,}{{Wang}
  et~al.}{2009}]{2009ApJ...699L.139W}
{Wang} X.,  et~al., 2009, \mn@doi [\apjl] {10.1088/0004-637X/699/2/L139}, \href
  {https://ui.adsabs.harvard.edu/abs/2009ApJ...699L.139W} {699, L139}

\bibitem[\protect\citeauthoryear{{Wood-Vasey}, {Wang}  \&
  {Aldering}}{{Wood-Vasey} et~al.}{2004}]{2004ApJ...616..339W}
{Wood-Vasey} W.~M.,  {Wang} L.,   {Aldering} G.,  2004, \mn@doi [\apj]
  {10.1086/424826}, \href
  {https://ui.adsabs.harvard.edu/abs/2004ApJ...616..339W} {616, 339}

\bibitem[\protect\citeauthoryear{{Yao} et~al.,}{{Yao}
  et~al.}{2019}]{2019ApJ...886..152Y}
{Yao} Y.,  et~al., 2019, \mn@doi [\apj] {10.3847/1538-4357/ab4cf5}, \href
  {https://ui.adsabs.harvard.edu/abs/2019ApJ...886..152Y} {886, 152}

\bibitem[\protect\citeauthoryear{{de Jong} et~al.,}{{de Jong}
  et~al.}{2019}]{2019Msngr.175....3D}
{de Jong} R.~S.,  et~al., 2019, \mn@doi [The Messenger]
  {10.18727/0722-6691/5117}, \href
  {https://ui.adsabs.harvard.edu/abs/2019Msngr.175....3D} {175, 3}

\bibitem[\protect\citeauthoryear{{van der Walt}, {Colbert}  \&
  {Varoquaux}}{{van der Walt} et~al.}{2011}]{2011CSE....13b..22V}
{van der Walt} S.,  {Colbert} S.~C.,   {Varoquaux} G.,  2011, \mn@doi
  [Computing in Science and Engineering] {10.1109/MCSE.2011.37}, \href
  {https://ui.adsabs.harvard.edu/abs/2011CSE....13b..22V} {13, 22}

\makeatother
\end{thebibliography}




\appendix
\section{Photometry pipeline description}
\label{sec:pipe_desc}
In this section, we present a detailed description of the photometric pipeline parameters. As described in section~\ref{ssec:phot_pipe}, the pipeline can be divided into three broad components, namely, reference building, image subtraction and forced photometry. Here, we describe each individual process.

\subsection{Reference building}
Since the SNe~Ia in our sample were discovered in 2018, in a fraction of the cases, the references for each field and readout channel were created during the lifetime of the SN, which manifests itself as non-zero flux significantly before and after the inferred time of maximum. For example, some SNe~Ia show negative flux in their forced photometry lightcurves before -30 days from the SALT2 $t_0$ time, since there is SN flux in the reference image. In our custom pipeline, we use images at least 30 days before peak to make deep references. For SNe with insufficient data ($< 30$ frames) in the phase range before -30 days, we also use data from $> 400$ days after peak. For each reference image, we have quality control selection cuts such that only images within the seeing range 1.7$^{"}$ and 3$^{"}$ and a nightly magnitude limit fainter than 19.2 mag are used in the final reference creation.

We use \texttt{SWARP} on the processed images from IPAC to make the reference. For each input image, we create a test weight map with the mask image from IPAC as an input \citep{2019PASP..131a8003M}. The weight map is passed as an to \texttt{SExtractor} to create an rms map for each input image that goes into making the final reference image. Each associated weight map, derived from the \texttt{SExtractor} output rms map, is input along with the image to make the reference coadd. Before coadding, the input images are all scaled to a common zero-point magnitude of 25.
The associated reference mask is produced by using the logical AND function with \texttt{SWARP} to combine the individual epoch masks. This is because logical OR coaddition is extremely restrictive at masking the reference image.

\subsection{Image Subtraction}
The individual science frame, and the associated mask and rms file, along with the reference coadd and its associated mask and rms file are input to the image subtraction routine. As described in section~\ref{ssec:phot_pipe}, we use \texttt{HOTPANTS} \citep{2015ascl.soft04004B} for the difference imaging part of the pipeline.  For both the science image and the reference, we fix the lower valid data count as the computed background minus 10 times the standard deviation. For the differencing, we use a convolution kernel with a half-width that is 2.5 times the seeing and a half width of the substamp to extract around each centroid as 6 times the seeing. We set the normalisation to the science image and convolve to the reference image. 

\subsection{Forced photometry}
We use the astropy package \texttt{photutils} \citep{Bradley_2019_2533376} to perform aperture photometry. We define the aperture using the \texttt{SkyCircularAperture} function with a 6 pixel diameter aperture. The location is defined as the median coordinate in each filter, derived from the alert packets \citep[see][for details]{fremling2020}. We apply a correction for a six pixel aperture to the final zero point for the observed fluxes, as provided in the IPAC data products. 



\bsp	
\label{lastpage}
\end{document}